\documentclass{iopjournal}
\usepackage{lmodern}

\usepackage{soul}
\usepackage{amsthm}

\usepackage[most]{tcolorbox}
\newtcolorbox{sectionsummary}[1]{
  colback=gray!5,
  colframe=black!50,
  boxrule=0.5pt,
  arc=2pt,
  left=6pt,
  right=6pt,
  top=6pt,
  bottom=6pt,
  fonttitle=\bfseries,
  coltitle=black,  
  title=#1
}

\theoremstyle{definition}
\newtheorem{definition}{Definition}

\makeatletter
\renewenvironment{thebibliography}[1]
     {\section*{\refname
        \@mkboth{\MakeUppercase\refname}{\MakeUppercase\refname}}%
      \list{\@biblabel{\arabic{enumi}}}{%
        \settowidth\labelwidth{\@biblabel{#1}}%
        \leftmargin\labelwidth
        \advance\leftmargin\labelsep
        \itemsep 0pt % �� elimina el espacio entre items
        \parsep 0pt  % �� elimina espacio adicional
        \usecounter{enumi}}%
      \sloppy\clubpenalty4000\widowpenalty4000%
      \sfcode`\.=\@m}
     {\def\@noitemerr
       {\@latex@warning{Empty `thebibliography' environment}}%
      \endlist}
\makeatother

\usepackage{amsmath,amsthm,amsfonts,amssymb}
\usepackage{booktabs}
\usepackage{graphicx}
\usepackage{fancyhdr}
\usepackage{float}
\usepackage{anysize}
\usepackage{titlesec}
\usepackage{enumerate}
\usepackage{xcolor}
\usepackage{bbold}
\usepackage{braket}
\definecolor{lime}{HTML}{A6CE39}
\DeclareRobustCommand{\orcidicon}{%
	\begin{tikzpicture}
	\draw[lime, fill=lime] (0,0) 
	circle [radius=0.16] 
	node[white] {{\fontfamily{qag}\selectfont \tiny ID}};
	\draw[white, fill=white] (-0.0625,0.095) 
	circle [radius=0.007];
	\end{tikzpicture}
	\hspace{-2mm}
}

\foreach \x in {A, ..., Z}{%
	\expandafter\xdef\csname orcid\x\endcsname{\noexpand\href{https://orcid.org/\csname orcidauthor\x\endcsname}{\noexpand\orcidicon}}
}

\begin{document}

%\articletype{Article type}

\title{Benchmarking Quantum Computers: Towards a Standard Performance Evaluation Approach}

\author{Arturo Acuaviva$^1$\orcidA{}, David Aguirre$^{1,2}$\orcidB{}, Rub\'en Pe\~na$^{2,*}$\orcidC{} and Mikel Sanz$^{1,2,3,4}$\orcidD{}}

\affil{$^1$Department of Physical Chemistry, University of the Basque Country UPV/EHU, Apartado 644, 48080 Bilbao, Spain}

\affil{$^2$BCAM - Basque Center for Applied Mathematics, Mazarredo, 14 E48009 Bilbao, Basque Country – Spain}

\affil{$^3$EHU Quantum Center, University of the Basque Country UPV/EHU, Apartado 644, 48080 Bilbao, Spain}

\affil{$^4$IKERBASQUE, Basque Foundation for Science, Plaza Euskadi 5, 48009, Bilbao, Spain}

\affil{$^*$Author to whom any correspondence should be addressed.}

\email{rpena@bcamath.org}


\begin{abstract}
The technological development of increasingly larger quantum processors on different quantum platforms raises the problem of how to fairly compare their performance, known as quantum benchmarking of quantum processors. This is a challenge that computer scientists have already faced when comparing classical processors, leading to the development of various mathematical tools to address it, but also to the identification of the limits of this problem. In this work, we briefly review the most important aspects of both classical processor benchmarks and the metrics comprising them, providing precise definitions and analyzing the quality attributes that they should exhibit. Subsequently, we analyze the intrinsic properties that characterize the paradigm of quantum computing and hinder the naive transfer of strategies from classical benchmarking. However, we can still leverage some of the lessons learned such as the quality attributes of a \textit{good} benchmark such as relevance, reproducibility, fairness, verifiability and usability. Additionally, we review some of the most important metrics and benchmarks for quantum processors proposed in the literature, assessing what quality attributes they fulfill. Finally, we propose general guidelines for quantum benchmarking. These guidelines aim to pave the way for establishing a roadmap towards standardizing the performance evaluation of quantum devices, ultimately leading to the creation of an organization akin to the Standard Performance Evaluation Corporation (SPEC).

\end{abstract}

\section{Introduction}
Proper performance analysis of devices is key to providing information to potentially interested users, allowing fair comparisons among different devices. Additionally, manufacturers use these insights to keep improving their systems. The measurement of performance for classical computer systems has traditionally been controversial. Historically, it has not been easy to reach an agreement in the field on seemingly simple concepts such as what metrics are suitable to assess performance, the workload that should be used for stress tests, and even what performance means. Despite this, researchers were able to continuously improve the ability of computers to solve increasingly complex problems using an intuitive approach. This early success partially contributed to the lack of rigour in performance evaluation, and the delay of its formal introduction. Fortunately, some ground rules were set by some preliminary works on benchmarking \cite{6448963, HowNotToLieWithStatistics, Jacob1995NotesOC,CharacterizingComputerPerformanceWithASingleDigit}. These rules trained researchers in data aggregation using the most appropriate measures of central tendency, as well as they depicted common pitfalls when designing test suites. Moreover, they paved the way for the introduction of standardization through Standard Performance Evaluation Corporation (SPEC) and Transaction Processing Performance Council (TPC) \cite{HowToBuildABenchmark,Eigenmann2001nl}. Eventually, 
authors were able to compile a solid framework for the evaluation of performance for classical computers, offering a systematic approach to benchmarking \cite{Lilja2005o,Jain1991-xw,Kounev2020mh,leboudec2010performance,Feitelson2015}.

The adoption of standardized benchmarks within a scientific community has been shown to have a positive impact on the overall scientific maturity of a discipline \cite{Sim_BenchmarkingToAdvanceResearch}. Benchmarks can help to establish standards and best practices for research and development, as well as provide a basis for objective comparison between different systems. This can lead to more rigorous and reliable research practices, and ultimately advance the field as a whole. Nowadays, the quantum computing community faces similar challenges in measuring performance as those initially encountered by classical computing researchers. Even more, quantum computing is a nascent field, with numerous technology proposals and emerging issues further complicating the situation.

Although quantum computers have not yet demonstrated a practical advantage over classical systems in real-world applications, this does not preclude the development of meaningful benchmarks. Moreover, there is theoretical evidence that quantum computers can offer significant advantages for certain problems \cite{Shor1994}. An analogous situation occurred during the early years of classical computing. During the early development of classical computing, benchmarks were primarily focused on evaluating low-level system, such as Whetstone \cite{WhetstoneBook} and Dhrystone \cite{DhrystonePaper}, rather than application-specific tasks. These benchmarks played a critical role in guiding hardware development and assessing system maturity.

In the early stages of classical computing, the lack  of standardized benchmarking rules allowed manufacturers to define their own benchmark, often leading to both unintentional and strategic biases. In some cases, manufacturers even aggressively optimized their compilers or CPUs to perform well on specific benchmarks \cite{Hennessy2012}. This trend contributed to the emergence of a ``benchmarking industry" driven more by marketing considerations than by objective performance evaluation.

The field of quantum computing is similarly vulnerable. The lack of standardized benchmarking rules, together with the increasing competition among major industry players vying to outperform their rivals in the quantum computing race, might lead to a situation that resonates with Goodhart's law \cite{goodhart1975problems}: ``Any observed statistical regularity will tend to collapse once pressure is placed upon it for control purposes". In other words, the absence of a rigorously formulated and scrutinized approach to benchmarking in quantum computing may distort research priorities and hinder the development of truly scalable and reliable quantum processors. This bias occurs due to benchmarking models playing a crucial role on favouring what portion of the total ``behavior space" is captured \cite{SkadronChallengesInComputerArchitectureEvaluation}. 

This is an already known challenge in the field of quantum computing \cite{osti_1775058,osti_1648785,osti_1639885,osti_1594677,quantumComputingSummitWhitepaperAtlanta}, and there are initiatives such as the P7131 Project Authorization Request (PAR) proposal from the IEEE to standarize quantum computing performance hardware and software benchmarking \cite{StandrdForQuantumComputingPerformanceIEEE}, or the proposals from IBM to define key characteristics of a quantum benchmark \cite{amico2023defining}. Still, again, as the Goodhart's law warned us, and the classic article from Dongarra {\it et al.} stated \cite{BlumeKohout2019,6448963}: ``Although benchmarks are essential in performance evaluation, simple-minded application of them can produce misleading results. In fact, bad benchmarking can be worse than no benchmarking at all." Therefore, we believe there is a clear need for a standardized benchmarking framework in the field of quantum computing to be able to assess how good the different attempts are at measuring performance.

In this article, we propose general guidelines to steer practitioners in the development of fair and good quantum benchmarks and drive quantum processor development in the right direction. Our methodology focuses on incorporating some lessons learned over decades from classical computer benchmarking theory, adapting them to the intrinsic features, limitations, and potentials of quantum computing. In this sense, we analyze the fundamental elements of both classical benchmarks and the metrics that comprise them, identifying the quality attributes they must exhibit to be considered good. We evaluate the main quantum metrics and benchmarks proposed in the literature to determine which satisfy these quality attributes. Additionally, we introduce precise definitions for commonly used terms in quantum benchmarking to ensure proper use of terminology. Finally, we introduce a benchmarking structure that provides an overview of the most important factors to be considered in benchmark creation. Our efforts aim to advance towards the standardization of the evaluation of quantum device performance.

This article is organized as follows. In Sec. \ref{Classical Benchmarking}, we provide an overview of the key elements of both classical processor benchmarks and metrics, emphasizing the fundamental properties they must exhibit to be considered good benchmarks and metrics. Additionally, we establish a common definition of various concepts used in the field of classical benchmarking. Finally, we review some examples of classical benchmarks and their limitations. Then, in Sec. \ref{Quantum Benchmarking}, we introduce some of the distinctive features of quantum computing devices that distinguish them from classical counterparts. We unify and define concepts to be used in the field of quantum benchmarking based on the definition of their classical counterpart. Furthermore, we review existing proposals for assessing the performance of a quantum computer, presenting the most relevant techniques, metrics, and benchmarks, and assessing which quality attributes these metrics and benchmarks fulfill. In Sec. \ref{RoadmapQB}, we propose general guidelines for any practitioner to follow when proposing a quantum benchmark. We also compile a series of open questions that need resolution on the path toward standardizing the comparison of quantum processor performance. Subsequently, we propose the creation of an organization dedicated to Standard Performance Evaluation for Quantum Computers (SPEQC). Finally, in Sec. \ref{Conclusions}, we present our concluding remarks.

\section{What can we learn from Classical benchmarking?}

\label{Classical Benchmarking}
Originally, the word \textit{benchmark} used to refer to ``a mark on a workbench used to compare the lengths of pieces so as to determine whether one was longer or shorter than desired" \cite{specglossary}.  Even though the term has clearly evolved, it retains the intuitive idea of comparing one object against another one to measure its properties. Hence, it is not a surprise that the concept of computer benchmark \cite{PP} is based on this very same idea, and it is usually defined as ``a test, or set of tests, designed to compare the performance of one computer system against the performance of others" \cite{specglossary}.

A natural question that then arises is what we mean by \textit{performance}. Performance is typically conceived in the classical computing scope as the useful work that a computer can get done in a given amount of time, and/or with a set of given resources. Indeed, in the past, it was common to hold the premise that ``the time required to perform a specified amount of computation is the ultimate measure of computer performance" \cite{CharacterizingComputerPerformanceWithASingleDigit}. 
For classical computers, there have been historically two major approaches to the performance analysis problem: (1) Experimenting and measuring the actual system, and (2) Experimenting with a model of the system. For the latter, for any kind of performance evaluation not limited to computing devices you can use a physical model or a mathematical model. For computational performance assessment, we prefer to build models that can represent the system in terms of quantitative and logical relations. These models can later be either solved using mathematical tooling to obtain an analytical solution, or simulated \cite{Zeigler2018fg,SimulationModelingAndAnalysisLawAverill}. Even though in this work we will focus on the measurements of experiments with the actual systems, it is possible that we will see other approaches involving system modeling in the future. In fact, model-based approaches have been used by the quantum community for the assessment of quantum networks \cite{Wu2021,Coopmans2021} and the characterization of both incoherent and coherent errors in quantum gates \cite{Tripathi}. Furthermore, other event-based models of quantum phenomena have also been published \cite{DeRaedt2005,Willsch2020}.

The aforementioned definitions already illustrate some of the inherent complexity of benchmarking. Besides these, the multidimensional nature of modern computational systems only exacerbates the problem. For instance, some systems may exhibit greater performance when constrained to a small set of resources, while others may perform better when given a longer total execution time. Also, the tests themselves must be appropriately designed and specified to ensure that they do not introduce performance biases. As an example, a particular test program may heavily rely on floating-point operations, while another may focus on intensive integer arithmetic, leading to performance biases that favor certain architectures over others on modern computer devices.

In order to accurately evaluate the performance of devices, it is important to consider their various characteristics. Based on past attempts, it has become evident that a comprehensive assessment requires a holistic approach that takes into account multiple factors \cite{CharacterizingComputerPerformanceWithASingleDigit}. Furthermore, the tests used in such assessments must meet certain standards of fairness to ensure that the results are not biased. Therefore, a rigorous and well-defined methodology, that is specifically tailored to the device being evaluated, is required. Failure to do so could result in inaccurate or misleading conclusions. 

In this section, we will provide an overview of the fundamental characteristics that a quality benchmark should exhibit. This will be followed by a discussion of the different types of benchmarks that can be developed. Additionally, we will address the properties that a \textit{good} metric should exhibit for benchmarking purposes. Furthermore, we will define relevant concepts used in the field of classical benchmarking. Finally, we will examine various instances of classical benchmarks and their associated constraints.

\subsection{Characteristics of a \textit{good} benchmark}
\label{GoodBenchmark}

Before delving into the characteristics that define a \textit{good} device/processor benchmark, and the importance of considering these factors when creating and using them, it is important to establish a common definition of the term. Although there are various interpretations of the term, we will agree with the definition of the term proposed by the Standard Performance Evaluation Corporation (SPEC) \cite{specglossary}:

\begin{definition}[Classical Benchmark]
A benchmark is a test, or set of tests, designed to compare the performance of one computer system against the performance of others.
\end{definition}

This terminology has been widely accepted in the field and provides a comprehensive definition of what a benchmark should involve. It is worth noting that other sources have also proposed alternative definitions for the term \cite{Lilja2005o,SkadronChallengesInComputerArchitectureEvaluation,Kounev2020mh,HowToBuildABenchmark,leboudec2010performance,Jain1991-xw}. We would like to remark that a benchmark is chosen in order to test the performance with respect to a given test suite, i.e. an application or set of applications. A better performance for this task does not necessarily mean a better performance for other tasks.

According to the definition, the crux of creating a high-quality benchmark rests in selecting appropriate tests that precisely reflect the types of workloads the system is expected to encounter, or that measure the anticipated behavior of the system during specific tasks relevant to the assessor. In other words, the tests should be representative of the performance of the system in real-world scenarios (i.e. the portion of the behavior space that is considered relevant) and should provide a realistic measure of the capabilities of the system related to the task. On classical benchmarking, researchers have been deeply studying how to model the workload so that it is representative of the behavior it models. Note that this definition implies that the workload is context-agnostic, but that is not the case. A workload is only a good fit for a problem when restricted to a certain context (e.g., imagine a workload given by a sequence of operations. If the operations can be executed in parallel by the device, or if the performance metric is not correlated to the devices capacity of solving operations, the workload might not be useful) \cite{Feitelson2015}. For classical devices, special detail has been paid to understanding the main contributing factors that need to be considered when modelling a workload, and multiple examples can be found in the literature \cite{ComputerResearchWorkload,ArtificialWorkloadDesign}. Indeed, in Ref.~\cite{Jain1991-xw}, R. Jain explicitly states that the discussion on selecting the most appropriate workload is the deepest pitfall into which a performance analysis project may fall. Alternatively, in a simplistic manner, we can state that the workload is ultimately the input for the tests, and hence the old computer science classic \textit{garbage in, garbage out} (GIGO) principle applies. 

Apart from characterizing the workload, the benchmark has to provide flexibility in both its implementation and usage to accommodate different hardware configurations, software environments, and use cases. This flexibility should extend to the ability to configure and customize the tests, as well as to the option of running them on a variety of systems, platforms, and architectures. By incorporating these features, a benchmark can provide a reliable and accurate measure of system performance and enable meaningful comparisons among different systems.

To be able to build a \textit{good} benchmark, we introduce below a list of quality traits to ensure that it is comprehensive, accurate, and representative of system performance. These attributes are not intended as strict definitions, but rather allow for a degree of flexibility, making them applicable across a wide range of contexts. This approach aligns with the benchmarking principles proposed by Kounev \cite{Kounev2020mh} and also reflect the most desirable characteristics of benchmarks proposed by other authors in the field \cite{HowToBuildABenchmark,GustafsonAndSnell_HINT,Henning_SpecCpu2000,Huppler_TheArtOfBuildingGoodBenchmark,Sim_BenchmarkingToAdvanceResearch}.

\begin{enumerate}
\item \textbf{Relevance}: A benchmark should closely correlate with behaviors that are of interest to users and provide relevant information about the aspects of performance of interest. Some examples are speed, scalability, accuracy, energy consumption, or quality.
\item \textbf{Reproducibility}: A benchmark should produce consistent results when run with the same test configuration, enabling users to replicate and verify results.
\item \textbf{Fairness}: A benchmark should allow for different test configurations to compete on their merits without introducing biases or artificial limitations.
\item \textbf{Verifiability}: A benchmark should provide confidence that its results are accurate and can be verified using well-established and transparent methodologies.
\item \textbf{Usability}: A benchmark should be easy to use and the cost of running it should be affordable in order to avoid roadblocks for users to run it. 
\end{enumerate}

An in-depth analysis of the different features presented above can also help us to understand how to better design benchmark models, and why they are important.

In particular, let us start by considering \textit{relevance}, the first key attribute of an ideal benchmark. Crafting a benchmark that correlates with the performance of the system on a specific domain poses a considerable challenge, as it often results in limited applicability beyond that particular area. Conversely, benchmarks designed with the intention of broader applicability tend to sacrifice the depth of meaningfulness in a specific scenario. Thus, achieving the optimal balance of relevance in benchmark design requires careful deliberation and trade-offs, ensuring that the benchmarks scope aligns with the intended application domain while maintaining a meaningful level of specificity \cite{Huppler_TheArtOfBuildingGoodBenchmark}. 

 Regarding {\it reproducibility}, we see that even though it may initially appear to be a straightforward feature to attain, it often proves to be hard to meet in practice. A system or process is deemed to be in a steady state when the variables governing its behavior remain constant over time and benchmarks typically operate in a steady state. We can model this definition by assuming that the properties of the workload/test of the benchmark are random variables $X_{i_1}, \dots, X_{i_n}$ on given instants $i_1, i_2, \ldots, i_n$. Then, a system is said to be in a steady state (stationary state) whenever the joint distribution $X_{i_1 + s}, \dots, X_{i_n + s}$, where $s$ is a shifting index, matches the original one. By taking $n=1$ and $n=2$, it is possible to prove that the properties must have the same expected value, variance, and covariance. To model process in which the properties are independent from each other, we can relax the condition to define \textit{weak stationary} states. This can be achieved by only enforcing $\mathbb{E}[X_i] = k$ (all properties have the same expected value), and the covariance of $X_i$ and $X_j$ for any $i,j$ is a function satisfying that $\text{Cov}(X_i, X_j) = \gamma (|i - j|)$. Either steady-state/stationary or weak stationary are enforced as prerequisites to achieve meaningful results in the performance analysis. 

Typical applications of the devices translate into variations in load due to factors such as user usage patterns. This disparity between benchmarking and real-world applications underscores the difficulty in achieving full reproducibility. The steady-state problem impacts other factors too such as fairness, by potentially introducing fluctuating biases. In general, replicating the dynamic conditions of actual usage becomes complex. As we will see in the next sections, this aspect becomes even more pertinent when considering quantum devices that lack substantial mitigation and error correction techniques, requiring frequent recalibration. A simplistic model of this non-stationary behavior for each property of the workload could be $X_i = C_{i\, \text{mod}\, T} + Z_i$ where $Z$ is the error component, $C$ is the cyclic part (defined as $C_t$ for $0 \leq t \leq T$), and $T$ is the period or cycle time for the calibration. 

As we have introduced, {\it fairness} can be substantially influenced by the conditions before executing the benchmark. Even more, fairness requires ensuring that multiple configuration rules can be applied when running tests while ensuring a common ground is established to allow comparisons. To facilitate fair comparisons, on classical computing a mechanism known as \textit{peak} and \textit{base} metrics were introduced by SPEC CPU2017 \cite{cpu2017}. \textit{Base} metrics require that the test programs, constituting the building modules of a test suite in a given language, are compiled using the same rules (i.e. compilation flags), in the same order. Also, these metrics are mandatory for the generation of the benchmark report. On the other hand, optional \textit{peak} metrics can be included allowing for greater flexibility, enabling the usage of different compilation rules for each benchmark, as well as feedback-directed optimization. 

Similarly, {\it verifiability} presents comparable challenges that we should carefully consider too. In industry, benchmarks are typically run by vendors who have a vested interest in the outcomes. In contrast, in academia results are subjected to a peer review process to be repeated and built upon by other researchers. \textit{Good} benchmarks should introduce \textit{self-validation}, since this may improve verifiability and usability, ensuring that results can be independently corroborated and effectively utilized. 

Finally, {\it usability} is an important feature that a benchmark should satisfy. When a benchmark is easy to use, it is more likely to be adopted by users, organizations, and industry. 

Even a meticulously designed benchmark incorporating the aforementioned key features is susceptible to manipulation by different agents with conflicting goals. As an illustrative example, in Ref.~\cite{Jain1991-xw}, R. Jain presents a thought-provoking discussion on strategies employed to gain an advantage in what he terms the \textit{ratio game}. A ratio game is a technique that involves using ratios with incomparable bases and skillfully combining them to benefit oneself. This is a common scenario for benchmarking, whenever we want to assess performance among distinct devices, normalizing against the performance of a reference device. For this, even when the geometric mean is properly used to average the normalized values \cite{HowNotToLieWithStatistics}, it is possible to derive rules for beating the ratio game. Upon analysis, three rules emerge regarding the ratio game:
\begin{enumerate}
    \item \textit{If a system is better on some benchmarks and worse on others, contradicting conclusions can be drawn in some cases}.
    \item \textit{If one system outperforms all others across all benchmarks, the application of any ratio game technique fails to yield contradictory conclusions}.
    \item \textit{Even when one system exhibits better performance than another on all benchmarks, the selection of an appropriate base {\rm \cite{P}} can be manipulated to showcase a more favorable relative performance}.
\end{enumerate}

These statements highlight the susceptibility of benchmarks to manipulation, emphasizing the need for cautious interpretation and critical evaluation when drawing conclusions based on the results from ratio games. Other researchers such as Mashey have also call out the difficulties of aggregating data, and the controversy of using one Pythagorean mean over the others, as well as stated the common agreement that ``performance is not a single number" \cite{Mashey2004}. Nonetheless, even though there is agreement on the fact that a single number is delusional, there are still discussions to determine what is the preferred metric, sparking debates over which numerical representations are deemed more optimal in each context. 

Finally, it is important to remember that while a \textit{good} benchmark can provide a reliable and accurate measure of system performance, it cannot solve the problem of evaluating conflicting aspects of a computer device. Instead, it can help to properly balance the trade-offs for a particular objective and influence its strengths and weaknesses. It is not possible for a single benchmark to excel in all areas, and there is no \textit{silver bullet} to solve this problem \cite{SkadronChallengesInComputerArchitectureEvaluation}. Therefore, multiple workloads and benchmarks are necessary to comprehensively evaluate the performance of a system. Each benchmark may focus on different aspects of performance and may use different metrics and methodologies. At best, we can obtain a set of benchmarks that will be representative of the behavior space we want to model and measure. By combining the results of these multiple benchmarks, it would be possible to obtain a comprehensive and accurate picture of the system. Ultimately, the choice of benchmark(s) should depend on the specific use case and the objectives of the evaluation, as the relevance requirement already stated.

\subsection{What types of benchmarks are there?} \label{subsection.TypesOfBenchmarks}

Benchmarks can be classified according to multiple criteria, ranging from the objective of what portion of the behavior space they want to assess to the way the workload is defined.  All of these categories fall under one of the following three different strategies when trying to assess the performance \cite{Lilja2005o,Kounev2020mh}: 
\begin{itemize}
    \item \textbf{Fixed-work} workloads: This benchmarking strategy refers to the types of benchmarks that focus on measuring the amount of time required to perform a fixed amount of computational work, as seen in the case of the old classic Dhrystone and Whetstone benchmarks. These benchmarks compiled a list of programs to be executed, the former favouring floating point operations, while the latter focuses on integer performance analysis \cite{WhetstoneBook,DhrystonePaper}. 
    \item \textbf{Fixed-time} workloads: This benchmarking strategy refers to the types of benchmarks that focus on measuring the amount of computation performed within a fixed period, as seen in the case of the SLALOM benchmark. This benchmark runs an algorithm to calculate the radiosity, with the performance metric being the accuracy of the answer computed within a fixed time \cite{SLALOMBenchmark}. 

    \item \textbf{Variable-work} and \textbf{variable-time} workloads: This benchmarking strategy refers to the types of benchmarks that allow both the amount of computation performed and the time to vary. A great example is the HINT benchmark that aimed to find rational upper and lower bounds for $\int_{0}^{1} \frac{1-x}{1+x}. \,dx$. The assessment was later performed by calculating the quality improvements per second (QUIPS), i.e. dividing the quality of the solution by the execution time \cite{GustafsonAndSnell_HINT}.
\end{itemize} 

In general, the performance of a system depends on the performance of multiple components of the system that are utilized during a certain operation. The problem with using \textit{fixed-work} benchmarks is that they introduce an intrinsic performance bottleneck when measuring the increment of improving a single component of the system. Benchmarking models have been developed changing the workload type precisely to address this concern, and further discussions on overcoming issues with the \textit{fixed-work} via using \textit{fixed-time} benchmarks can be found in Ref. \cite{Gustafson1992, Isoefficiency}. However, benchmarks that allow for variability in both time and workload provide greater flexibility to evaluate the performance of different systems, determining which machine produces the highest-quality result in the shortest time possible.

One approach to benchmarking classification is obtained by focusing on the level of implementation required from users\cite{HowToBuildABenchmark,Kounev2020mh}: 
\begin{itemize}
    \item \textbf{Specification-based benchmarks} define a problem or functions that must be implemented, the required input parameters, and the expected outcomes. The responsibility for implementing the specification falls upon the individual conducting the benchmark. 
    \item \textbf{Kit-based benchmarks} include the implementation as a necessary part of the official benchmark execution.
    \item \textbf{Hybrid benchmarks} provide some functions as part of the kit, while others are allowed to be implemented at the discretion of the individual conducting the benchmark. 
\end{itemize}

The specification-based benchmarks are sometimes called ``pen and paper" benchmarks, and they have proved successful when there are considerable differences in the architectures on which the benchmarks will be implemented \cite{ReflectionsOnCreationParallelBenchmarkSuite}. Still, they come at the cost of ad-hoc and expensive implementations from their users, as well as challenges with verifiability. On the other hand, kit-based benchmarks reduce the cost and time to run them by offering execution-ready implementation, which favoured them historically.  Indeed, the stone aged benchmarks for classical computing are good examples of this, such as the Whestone, Dhrystone or the Linpack benchmarks \cite{WhetstoneBook,DhrystonePaper,OverviewOfCommonBenchmarks}. 

Another approach to benchmarking classification is obtained by focusing on the workload \cite{Kounev2020mh,Lilja2005o}:
\begin{itemize}
    \item \textbf{Synthetic benchmarks}, when artificial programs are constructed to try to mimic the underlying characteristics of a given class of applications.
    \item \textbf{Microbenchmarks}, small programs used to test specific parts of the system, independent of the rest (e.g. evaluate the I/O (input/output) subsystem in a computer).
    \item \textbf{Kernel benchmarks} or program kernels, small programs that capture the key portion of a specific type of application (e.g. execute the portion of a program that consumes a larger fraction of the total time within an application). 
    \item \textbf{Application benchmarks}, complete real programs designed to be representative of a particular class of applications. 
\end{itemize}

Finally, we must not forget that to be relevant, benchmarks must correlate with the expected behavior that wants to be assessed. Hence, there are different benchmarks that deal with different goals in classical computing. For instance, we could benchmark memories, GPU, Networks, Databases, Power Consumption, CPU-intensive operations, and each of them might require customised benchmarks to properly capture its behavior space.

\subsection{What makes a \textit{good} metric}
\label{GoodMetric}
As discussed, a benchmark is a test or workload that we characterise according to certain criteria, we must always later \textit{measure} the results of each test. The measurement of these results can range from the time it takes to finish to more complex metrics such as the quality of a solution. To further break down how to properly define these metrics, let us start first introducing the term:

\begin{definition}[Performance Metric] \label{Definition.PerformanceMetric}
    A value derived from the evaluation of a function that maps a considered set of objects or events to another set of values, characterizing a given performance-related property of a system.
\end{definition}

From Lilja and Kounev \cite{Lilja2005o,Kounev2020mh}, we know that these can be typically classified into three groups: Counting how often a certain event occurs (e.g. number of operations processed in a given observation interval), duration of an interval (e.g. time taken to complete a single operation), and size of a parameter (e.g. precision obtained to compute a value). Some examples of basis performance metrics are the response time (i.e., time taken by a system to react to a request providing a respective response), throughput (i.e., rate at which requests are processed by a system), and utilization (i.e., fraction of time in which a resource is used). 

Similarly to the quality attributes defined for a benchmark, there are properties proposed in the literature for an ideal performance metric \cite{Kounev2020mh,Lilja2005o,HowToBuildABenchmark}:

\begin{enumerate}
\label{enum:good_metric}
    \item \textbf{Practical}. A metric is practical if it can be measured in a viable and realistic time. This is complicated to define, but we will call so if it can be efficiently computed, i.e. within polynomial time in terms of the relevant size of the problem, thus facilitating its practical applicability. We understand that for some polynomials and sizes it can be challenging, but we must add a threshold. Additionally, this will help meet the \textit{usability} criteria for the benchmarks using it.
    \item \textbf{Repeatable}. If the metric is measured multiple times using the same experiment, the same value must be measured. Ideally, a metric should be deterministic when measured multiple times. In practice, small differences are acceptable, provided they are bounded. Note that this will allow the benchmark to meet the \textit{reproducibility} criteria.
    \item \textbf{Reliable}. It must rank systems consistently with respect to a property that is subject to evaluation. In other words, if system $A$ performs better than system $B$ with respect to the property under evaluation, then the values of the metric for the two systems should consistently indicate so. The reliability of a metric is closely correlated with the \textit{relevance} of a benchmark. 
    \item \textbf{Linear}. Linear metrics are intuitively appealing since humans typically tend to think in linear terms: twice as high value of the metric should indicate twice as good performance.
    \item \textbf{Consistent}. It has the same units and the same precise definition across different systems or configurations of the same system. When this property is not met, the benchmark will suffer poor \textit{verifiability} and potentially lack of \textit{relevance}. 
\end{enumerate}

Not all of the historically ideal properties of a metric must be preserved for a metric to be \textit{good enough} to build a \textit{good} benchmark. For instance, linearity is not preserved for the Ritcher scale but this does not prevent us from leveraging it as an useful metric. On the other hand, independent metrics could act as a sufficient but not necessary condition for the lack of influence on the metric to favour certain manufacturers. In general, when a metric does not fulfill a certain condition, it is actually possible to compensate it with other metrics within the same benchmark.

We can also classify metrics depending on what they measure. Metrics that measure what was done, useful or not, have been called means-based metrics whereas end-based metrics measure what is actually accomplished. As an example, for a given program we could measure the floating point operations per second (FLOPS), or the total execution time. The former is a mean-based metric, while the latter is an end-based metric. Depending on the context, it could be argued that one is better than the other \cite{Lilja2005o}. 

Other well-known examples of classical metrics are clock rate, or million instructions per second (MIPS). The clock rate is repeatable (given a fixed processor), practical, consistent, and independent metric. However, given its limited scope, the clock rate is both non-linear and unreliable as a performance metric (a processor with a faster clock rate does not necessarily imply better overall performance. The clock rate ignores many important performance-relevant aspects such as how much computation is actually performed in each clock cycle as well as the interaction with the memory and I/O subsystems). On the other hand, MIPS is neither reliable, linear, or consistent (for the latter, take the example of different Instruction Set Architectures, such as RISC vs CISC \cite{IsaWars}). Ultimately, we mentioned before FLOPS. Unfortunately, this metric is neither reliable, linear, consistent, nor independent (processor manufactures may use different rules for what is counted as floating-point operations, which makes the metric fail for consistency and independence). 

Finally, it is important to highlight two key points. First, the properties discussed in this work are applicable not only to full-stack benchmarks and metrics but also to individual components of the computing stack. Second, the seemingly straightforward task of measuring a property using a metric might be tough. The effect of the Simpsons' paradox serves as an adage to the designers of benchmarks and metrics, calling out the possibility of reversing or disappearing the trend on the data when combining it \cite{leboudec2010performance}.

\subsection{Examples of classical benchmarks}

Classical computing benchmarking has evolved to address multiple and distinct application domains, from machine learning to high-performance computing (HPC), each with domain-specific requirements. Experience with classical benchmarks has shown that defining benchmarks that remain relevant, fair, and consistent is a nontrivial challenge. For instance, the extensive use of LINPACK and its successor HPLinpack for ranking supercomputers has revealed that such benchmarks may fail to accurately reflect performance on real-world HPC applications, while also encouraging optimization strategies tailored to the benchmark rather than to relevant workloads \cite{dongarra1979linpack,Dongarra2003TheLB,top500,CriticsTop500,PerformanceModelingHPCG}. Similarly, synthetic benchmarks such as Whetstone \cite{WhetstoneBook} have proven controversial. Although designed to measure floating-point performance, they can be misleading, as modern architectures may exploit cache effects and pipeline optimizations that are unrepresentative of real workloads. Historically, this issue was further compounded by the lack of consistent definitions of what constituted a “floating-point operation” across architectures and instruction sets, making cross-platform comparisons ambiguous in early benchmarking efforts. More generally, synthetic benchmarks often correlate poorly with practical application performance, limiting their relevance. These limitations highlight the need for continuous adaptation of benchmarking methodologies as technology evolves. This approach has been adopted by the Standard Performance Evaluation Corporation (SPEC), which regularly updates its benchmark suites and develops application-specific benchmarks targeting different workload classes, such as SPECviewperf for graphics workloads, SPECaccel for accelerator-based applications, and SPEC CPU for compute-intensive workloads \cite{gwpg.spec.org,spec.org.accel2023,spec.org.cpu2017}. These lessons from classical benchmarking motivate a careful, principle-based approach to quantum benchmarking, especially given the rapid evolution and heterogeneity of quantum hardware platforms.

\begin{sectionsummary}{Key points of the section: What can we learn from Classical benchmarking?}
\begin{itemize}\setlength\itemsep{2pt}

    \item \textbf{What benchmarking is}: In computing, a benchmark is a test (or set of tests) designed to compare the performance of one system against others.

    \item \textbf{What performance means}: Performance is defined as the useful work done within a specific time or using specific resources.

\item \textbf{Properties of a ``Good'' Benchmark:} We adopt the SPEC definition, which outlines five essential qualities:
\begin{itemize}
    \item \textit{Relevance}: Correlates with user interests (speed, energy, accuracy).
   \item \textit{Reproducibility}: Produces consistent results under the same configuration. 
    \item \textit{Fairness}: Allows competition without bias (e.g., using ``Base'' metrics for standard compilation rules vs. ``Peak'' metrics for optimized flexibility).
    \item  \textit{Verifiability}: Results can be independently confirmed; ideally includes self-validation.
    \item \textit{Usability}: Easy and affordable to run.
\end{itemize}

\item \textbf{Types of Benchmarks:} Benchmarks are classified by three main criteria:
    \begin{itemize}
        \item By Strategy (Time vs. Work): Fixed-work, fixed-time, and variable-work/variable-time.
        \item By Implementation Level: Specification-based, kit-based, hybrid.
        \item By Workload type: Synthetic, microbenchmarks, kernels and applications.
    \end{itemize}
   
    %\item Fixed-work: Measures time to complete a set task (e.g., Dhrystone). Drawback: Can introduce bottlenecks.
   %\item  Fixed-time: Measures work done in a set period (e.g., SLALOM).
    %\item  Variable-work/Variable-time: Both vary to find quality improvements per second (e.g., HINT).
    %\item Specification-based ("Pen \& Paper"): User implements a described problem (hard to verify, high effort).
    
    %\item  Kit-based: Code is provided (e.g., LINPACK; easier, standard).
    
    %\item Hybrid: A mix of both.
    
    %\item Synthetic: Artificial programs mimicking application characteristics.

    %\item Microbenchmarks: Test specific components (e.g., I/O).

    %\item Kernel: Small, key portions of a larger application.

    %\item Application: Complete, real-world programs.
    %\end{itemize}

\item \textbf{Metric}: A metric characterizes a performance property (count, duration, or size), and can be classified in two categories: means-based that measures what was done (e.g., FLOPS), and end-based that measures what was accomplished (e.g., total execution time). Following SPEC, a ``good'' metric should be:
    \begin{itemize}
        \item \textit{Practical}: Efficient/polynomial time to compute.
        \item \textit{Repeatable}: Deterministic results.
        \item \textit{Reliable}: Consistently ranks systems (System A $>$ System B).
        \item \textit{Linear}: Twice the value equals twice the performance.
        \item \textit{Consistent}: Same units and definitions across systems.
   \end{itemize}

\item \textbf{Lessons from History}
    \begin{itemize}
        \item Adaptation is Key: Benchmarks must evolve. LINPACK is used for ranking supercomputers but often fails to reflect real-world HPC applications, encouraging optimization for the benchmark rather than the workload.
    
        \item Synthetic Risks: Benchmarks like Whetstone can be misleading on modern hardware because they do not account for modern cache effects or pipeline optimizations.
    
        \item No Silver Bullet: No single benchmark can evaluate all aspects of a computer. A comprehensive assessment requires a suite of benchmarks (like SPEC CPU or SPECviewperf) to model the full behavior space.
    \end{itemize}
    
\end{itemize}
\end{sectionsummary}

\section{Quantum benchmarking}
\label{Quantum Benchmarking}
Quantum computers are devices that leverage the principles of quantum mechanics to address challenges beyond the computational capacity of classical computers. Several physical implementations for quantum computers have been explored, each employing different technologies as qubits, such as trapped ion \cite{QCion,Zhang_2017}, superconducting circuits \cite{Arute2019,ZHU2022240,PhysRevLett.127.180501}, photonic \cite{QCPhoton,PhysRevLett.127.180502}, and cold atoms \cite{Bernien2017ProbingMD,Ebadi2020QuantumPO}. Each of these architectures has its own set of advantages and disadvantages, and research is ongoing to determine which approach is most suitable for scaling up quantum computers to a practical level. Additionally, various quantum computing paradigms have emerged, including annealing-based quantum computing, which relies on the adiabatic theorem to solve computational problems \cite{farhi2000}; analog quantum computing, in which a controllable quantum device used to simulate other quantum systems \cite{Kendon_2010}; digital quantum computing, which relies on the quantum circuit model for implementing general quantum algorithms \cite{doi:10.1126/science.273.5278.1073}; and measurement-based quantum computing, where computation is performed through a sequence of measurements on a highly entangled state \cite{Raussendorf}. These four paradigms offer different perspectives on the practical applications of quantum computing. 

The diversity of technological platforms and paradigms used in the development of quantum computers introduces significant complexity in comparing the performance among different devices. For this reason, the search for a common benchmark that allows comparison of quantum computers performance across different technological platforms and paradigms poses a challenge to the scientific community. Despite these challenges, efforts have been made to develop benchmarks enabling comparison among diverse quantum computers. Primarily, these efforts focus on three types of relevant comparisons: comparing the performance of different quantum processors within the same computational paradigm (e.g., comparison of digital quantum computing platform providers such as Forest, Qiskit, ProjectQ \cite{LaRose2019overviewcomparison}, or comparisons between quantum processors \cite{Linke2017,miessen2024benchmarkingdigitalquantumsimulations,Ilin_2024,wright2024noisyintermediatescalequantumsimulation}); comparing different computational paradigms (e.g., between processors for digital quantum computing and quantum annealing or digital-analog quantum computing \cite{Flannigan_2022}); and finally, comparing quantum computers to classical computers (e.g. comparing digital processing against quantum annealing \cite{Jnger2021}, empirical comparisons relying upon simplified Machine Learning (ML) tasks \cite{havenstein2018comparisons,Cugini2023} or even the claims on quantum supremacy \cite{Arute2019,kalai2022google}). It is worth noting that these three comparisons are crucial for advancing scalable and reliable quantum technologies, but each of them demands a thorough analysis.  Furthermore, a similar analysis could be expanded to include algorithmic comparisons. Looking ahead, benchmarking should be extended to cover other potential new aspects of a quantum computer that we have not even envisioned yet (e.g. qRAM \cite{QuantumRam} or other future components that we have not even thought about yet). During this work, we will restrict ourselves to the evaluation of the performance of quantum processors.

Despite technological advances, we are still undergoing the development of the quantum computing foundations. As Blume-Kohout and Young stated \cite{osti_1594677}, we are still more in the early stages of the quantum computing technology world rather than close to any exponential growth, which translates into challenges as we do not have clearly defined applications for these devices. What is more, we lack knowledge of how each the performance of each component will propagate and how the different quantum technology architectures will evolve. The current state is closely related to classical devices in the pre-1940 stages with the introduction of devices that were able to mechanize sequential series of calculations and procedures in binary code via the Bell Labs Model I, which served as an example of successfully applying Boolean logic to the design of computers \cite{TROPP1980115}. Back then, computers were starting to be explored, in contrast to upcoming years such as 1965 when Moore published the famous paper that led to the establishment of the Moore's law \cite{MooreLaw}, and exponential growth took off. This contrasts with the ``benchmarketing" approach that we sometimes see in the community when promoting the results of quantum computing experiments. 

In the quantum computing field, the term benchmarking has been used in a variety of contexts, sometimes leading to ambiguity. While it traditionally refers to a series of tests evaluating the performance of a device, its application has expanded to encompass the metrics evaluated as part of the test, the actual tests, or even for verifying the success of a test. We believe that establishing a unified and precise terminology for quantum benchmarking is essential to enable consistent and effective evaluation of quantum systems.

In this section, we introduce some of the intrinsic characteristics of quantum computer devices that make them different from classical ones, making the benchmarking model non-transferable. Additionally, we review the technological stages that we expect these devices will undergo throughout their evolution. Subsequently, we propose a unified terminology for the quantum benchmarking field. Finally, we review some existing proposals metrics and benchmarks for quantum processor in the literature, assessing what quality attributes they satisfy.

\subsection{Nuances of benchmarking quantum computers}
\label{subsection:nuances of QC}

There is lack of a definite computational paradigm, standard, platform, or even architecture for quantum computers. This is an eventual consequence of our limited knowledge so far about the real scope of quantum computing devices and platforms. Note that, in classical computing, regardless of the type of architecture (i.e. Princeton/Von Neumann vs. Harvard), there is a similar unified approach that allows for performance evaluation using well-established metrics, workloads, and even benchmarks. Indeed, it can be argued that differences in the architectures have been historically deceptive \cite{mythHarvard}, and that devices can be easily compared assuming a similar common ground. Interestingly, this was not always the case. As mentioned before, back in the times of the Bell Labs Model I, every computer device was custom and hard to compare against each other. Benchmarking was not so extended and generalised as a recommended best practice until the 70s-80s. 

Similar to these early years, quantum computers are a field in its first stages. Based on our current expectations, we could clearly divide the future different quantum technological development eras as it follows:

\begin{enumerate}
    \item \textbf{Noisy Intermediate-Scale Quantum (NISQ)} \cite{Preskill2018},
    characterized by middle-size quantum processors with tens to hundreds of
    noisy physical qubits, substantially limiting the size of the quantum
    circuits that can be reliably executed. In the current state of NISQ era, it is controversial whether quantum processors have already reached the necessary technological development to achieve quantum advantage.
    
    \item \textbf{Partially Quantum Error Corrected (PQEC)}, when quantum
    processors errors are below a certain threshold that allows the application
    of error correction codes, such as the surface code \cite{Fowler2012}. When
    an insufficient number of physical qubits are available, we could employ a
    smaller code and partially correct errors up to a certain distance-code and
    number of errors per per cycle of stabilizer measurements
    \cite{Andersen2020}.
    
    \item \textbf{Fault Tolerant Quantum Computers (FTQC)}. Once quantum error correction codes can be efficiently implemented, it will be possible to run deeper and more complex quantum circuits without errors spreading uncontrollably. Although errors will not disappear completely, quantum information will be protected from the environment through correction schemes that limit the accumulation and spread of errors.
    
    \item \textbf{Fully Functional Quantum Computers (FFQC)}. Existing quantum
    algorithms and circuits consider the use of ancillary qubits, which play a
    similar role as  cache memory in classical processors. However, for Fully
    Functional Quantum Computers, there will also be necessary the existence of
    Quantum Random Access Memories (qRAM) \cite{QuantumRam}, and even long-term
    (error-corrected) quantum memories \cite{Terhal2015}. 
\end{enumerate}

We live currently in the NISQ era, while the community is still investing to keep advancing quantum computers to the next stages. To meet the PQEC era, we need to further explore error correction codes. It is important to consider that the number of physical qubits required by the error corrections codes to correct any error generally depends on how below the threshold the processor and operations are \cite{Andersen2020,suppressingQuantumErrors}. The generation of quantum processors developed in this era could be considered a transition age towards a fault-tolerant quantum computer in which, in principle, there are no limitations in the size of the quantum circuits that can be reliably executed. It is important to remark that these codes rely on sufficiently trustworthy qubits and operations, not only quantum gates, but also very accurate measurements. 

The challenges that we are facing today within the NISQ era are different from the ones we will be facing in the upcoming eras, and hence the benchmarks should reflect this. Given the wide diversity of technological platforms and paradigms for implementing quantum computers, it is likely that benchmarks in the NISQ era will focus on comparing the performance among different quantum computers within subsets of the broad existing variety. As quantum computing technology advances, it is likely that certain technological platforms will become dominant, particularly those that demonstrate scalability, fault tolerance, or economic viability for large-scale implementation. However, this convergence does not avoid the possibility of continued diversification. As seen in classical computing with the emergence of application-specific integrated circuits (ASICs), technological maturity can also drive specialization tailored to distinct tasks. Similarly, in quantum computing, different platforms may evolve to address specific application domains. In Sec. \ref{RoadmapQB}, we will discuss further what recommendations we suggest to properly face the different quantum technological development eras.

As mentioned at the beginning of this article, quantum computing exhibits fundamental characteristics and limitations associated with the nature of quantum physics, rather than being tied to technological development. Consequently, classical benchmarks and metrics cannot be directly extended to quantum computing. We now summarize some of these intrinsic characteristics of quantum computing:
\begin{figure}[t!]
    \centering
    \includegraphics[width=0.70\columnwidth]{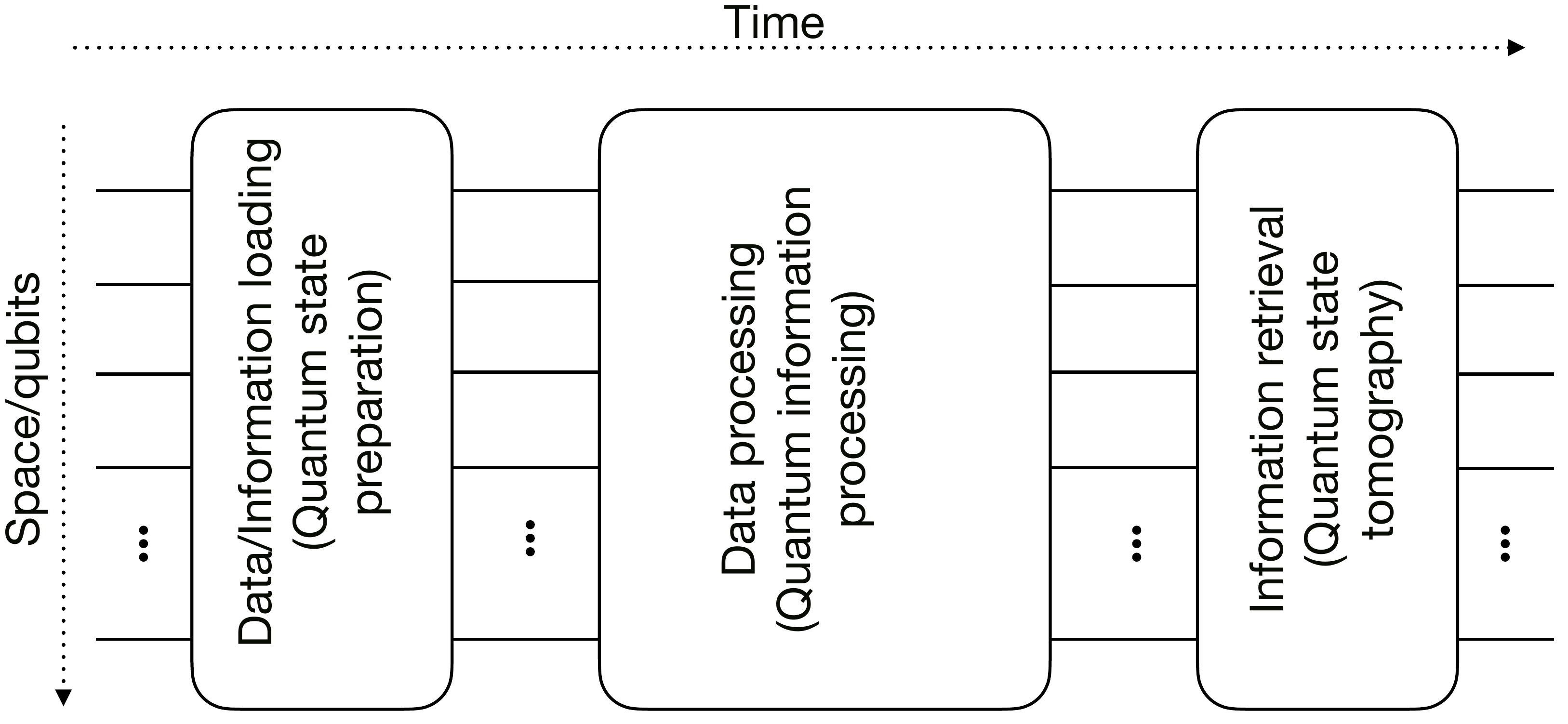}
    \caption{{\bf Input-Processing-Output (IPO) model}. IPO model can be applied to many quantum algorithms, with quantum state preparation as input, quantum information processing as data processing, and quantum state tomography as information retrieval. }
    \label{Fig:IPOmodel}
\end{figure}

\begin{itemize}
     \item The fundamental unit of information in a quantum computer is the qubit (quantum bit). While classical bits can only represent a single
     binary value 0 or 1, a qubit is described as a linear combination of the $|0\rangle$ and $|1\rangle$
     states.

    \item Even though an $n$-qubit quantum state can in principle contain an
    exponential quantity of information in bits \cite{10.1145/581771.581773},
    the amount of classical information that can be retrieved by a projective
    measurement is limited to, at most, $n$ bits due to Holevo's bound
    \cite{Maslov_2021}. As a consequence, if the algorithm requires to retrieve
    the full quantum state, i.e. to perform full tomography of the state, then
    an exponential number of measurements will be required.

    \item After every run of the algorithm, if the information is classically
    extracted, the processor must be restarted in the initial state. This state
    is typically represented as $|\psi_0\rangle=\otimes_{i=1}^n|0\rangle_i$,
    where $n$ denotes the number of qubits. This process is usually slow and has
    to be taken into account in the running time of the algorithm.

    \item A quantum computer execution typically sticks to the Input-Process-Output (IPO) model, a rather general framework used in software engineering to describe the structure of an information processing program. Generally, a quantum computer needs to complete three different stages: the input or loading of information, the processing thereof, and the output or retrieval of relevant information. Most quantum algorithms adhere to this model \cite{PPP}, where the input is state preparation or matrix exponentiation, and the output is just the quantum measurement (see Fig.\ref{Fig:IPOmodel}), adding extra steps to the classical algoritmic execution.

    \item On a quantum computer, after processing, we have a quantum state that needs to be measured in a basis. Measurement involves projecting the quantum state using a series of projectors that collapse the state to one of the bases (single-shot measurement). To obtain the probabilities of each state, this process must be averaged over $N$ repetitions, where $N$ depends on the precision with which we aim to recover the coefficients and the number of qubits comprising the state. More specifically, as every run is independent of the previous ones, the variance in the estimation is asymptotically given by the Cram\'er-Rao bound $(\Delta p)^2 \geq (F N)^{-1}$, with $F$ the Fisher information associated to the aforementioned projective measurement.

    \item While classical computers programs executions can be optimised by the compilers, quantum computers can be optimised during the transpilation process of the quantum circuit (e.g. translating a Qiskit circuit into OpenQASM) as well as during the translation of the circuit into the device custom control logic to operate the device (e.g. translating OpenQASM into microwave pulses for a superconductor quantum computer). This means that there are two level of abstractions that offer potential for optimizations, and therefore can impact the performance assessment of a device.

\end{itemize}

Beyond these fundamental features associated to quantum physics, there are other relevant aspects of quantum computing which mainly affect the performance for the NISQ and PQEC eras.

\begin{itemize}
    \item The number of qubits in quantum computing is a crucial factor that directly influences the capabilities and potential of quantum computers. Increased quantities of qubits empower the quantum computer to solve more complex problems. However, it is not only the quantity of qubits that is important, but also their quality.
    
    \item Qubit connectivity plays a key role in the performance of quantum computing because it determines the interactions between the qubits. To perform quantum computations, qubits need to interact through quantum gates. When qubits are not directly connected additional operations, referred to as swap operations, must be applied to facilitate their interaction. These additional operations increase the depth and runtime of the quantum algorithm, affecting the performance of noisy quantum computers, especially in NISQ and PQEC eras.

    \item While errors in classical computing are extensively studied and managed through classical error correction protocols, noise significantly affects the performance of quantum computing systems. Quantum computers are inherently susceptible to errors and inaccuracies due to various sources of noise such as thermal environment and external control factors. The effects of noise affect different aspects of quantum computation. High gate error rates reduce the fidelity of quantum operations, affecting the accuracy of computations. Furthermore, the interaction with an environment causes decoherence, leading to the loss of quantum information. High decoherence levels limit the circuit depth that can be accurately executed on the quantum computer, directly affecting the complexity of the algorithms that can be executed. Furthermore, NISQ devices require periodic recalibration due to the effect of the thermal environment and the external control.

\end{itemize}

The number of qubits available and their quality, the connectivity, and the noise will have a greater impact on NISQ and PECQ devices than on the later stages for the assessment of the performance. The routing and mapping of an algorithm will play a key role in the performance of a quantum computer, as simple changes on how the algorithms are executed can lead to drastically different results even for the same hardware.

\subsection{Why platform heterogeneity precludes rigid numerical standards}
\label{subsection:platform heterogeneity} 

In classical computing, heterogeneity among devices is mainly architectural, while the underlying operational model and logical abstractions remain consistent across platforms. In contrast, quantum computing is currently in a stage characterized by a high degree of platform diversity. Quantum platforms differ in several aspects, including the underlying physical technologies used to implement qubits, the set of native quantum operations they support, and the topology of the quantum hardware. As a consequence, performance-related properties such as gate execution times, qubit connectivity, and dominant error mechanisms exhibit differences across platforms, challenging the use of numerical thresholds to define performance acceptability. For this reason, specifying quantitative cutoffs for performance properties is inappropriate at this stage.

For instance, gate execution times span several orders of magnitude across platforms, ranging from microseconds to milliseconds, depending on the underlying physical technology. As a result, a metric that relies on the execution of deep quantum circuits may be feasible on platforms with fast gate operations, while becoming increasingly costly on platforms with significantly slower gates due to increased experimental runtime and accumulated decoherence. Consequently, defining what constitutes an acceptable or good execution time through a fixed numerical threshold does not carry the same meaning across different technologies. Likewise, quantum platforms differ in the native gate sets they support, ranging from architectures limited to one- and two-qubit gates to platforms enabling native multi-qubit operations. As a consequence, metrics based on gate counts or circuit depth may correspond to different computational effort across platforms, since a single native multi-qubit operation can replace an entire sequence of two-qubit gates in other architectures. Therefore, defining acceptable or meaningful thresholds on gate counts or circuit depth through fixed numerical values does not reflect equivalent operational cost across platforms. Similarly, qubit connectivity differs widely across platforms, ranging from nearest-neighbour connectivity to all-to-all connectivity. As a result, metrics that rely on interactions between distant qubits may incur different overheads depending on the connectivity of the quantum processor. Furthermore, error mechanisms vary qualitatively across technologies, with different platforms being dominated by distinct noise sources such as decoherence, control errors, crosstalk, or measurement noise. Therefore, metrics or benchmarks sensitive to particular error sources may reflect different performance constraints across platforms. As a result, fixed numerical thresholds on such metrics do not carry the same meaning across technologies.

These examples are only illustrative. In general, the heterogeneity of quantum platforms reinforces the difficulty of defining universal numerical thresholds for performance acceptability. Taken together, these considerations illustrate that the difficulty in defining rigid numerical standards for quantum benchmarking comes from the heterogeneity of quantum platforms. At the present stage of quantum hardware development, a guideline-based framework provides a more robust approach, allowing benchmarks and metrics to be evaluated in context without imposing premature numerical standardization.

\subsection{A common language to discuss}

The lack of a common and accepted standard for defining techniques and processes in quantum computing has led to the misuse of terminology. Even though there has been past attempts to classify the different types of benchmarks, and review approaches in the field \cite{hashim2024, lall2025reviewcollectionmetricsbenchmarks}, the terms quantum benchmarking, validation, verification, and testing have been loosely and inconsistently used in the literature. While we appreciate the groundwork laid by \cite{amico2023defining,proctor2024benchmarkingquantumcomputers} in establishing a foundation for benchmarking standards and properties, our approach builds upon not only insights from quantum computing research but also draws from the well-established and mature field of classical computing benchmarking. In addition, we support our proposal with numerous carefully analyzed examples, which help to illustrate and validate the concepts and classifications we introduce. In this subsection, we will propose a unified terminology for the main concepts used in the quantum benchmarking field.

To ensure alignment on terminology, we precisely define below some of the terms commonly used in quantum benchmarking, with definitions inspired by the classical ones:

\begin{definition}[Quantum Benchmark]
A test, or set of tests, that aims to measure or evaluate the performance, efficiency, or other properties of a quantum processor or hardware component for a certain task.
\end{definition}

The underlying goal of a quantum benchmark is to be able to compare devices with
each other. What is more, similarly to classical benchmarking, quantum
benchmarks involve running specific tests or workloads on the systems and
software collecting data to analyze its performance. Examples of quantum benchmarks include QASMBench \cite{li2022qasmbench}, Quantum Volume \cite{PhysRevA.100.032328}, or Quantum LINPACK \cite{QuantumLINPACK}.

When executing these benchmarks, we need to later compile a list of metrics that
will describe the system. We introduce the definition of a performance metric
analogous to the classical case:
\begin{definition}[Quantum Performance Metric]
    A quantum performance metric is a performance metric, Definition
    \ref{Definition.PerformanceMetric}, where the property belongs to a quantum
    system. 
\end{definition}

When the contextual information provides the details about the system being
assessed, we briefly refer to them as just metrics. Interestingly, Quantum
Volume is not only a benchmark but also a metric \cite{PhysRevA.100.032328},
even though it is tightly coupled to the Quantum Volume benchmark definition.
Other examples of metrics are cross-entropy difference, $\ell_1$-Norm Distance
to measure total divergence of probability distributions to the sample space,
the gates fidelity, the number of qubits, or the Q-Score
\cite{Martiel2021BenchmarkingQC}. 

Following is the definition of \textit{quantum verification}, which refer to the
process of certifying that a quantum device works as it is expected:

\begin{definition}[Quantum Verification] \label{Definition.QuantumVerification}
   Process of evaluating whether or not a quantum computer is indeed producing correct results.
\end{definition}

This concept is particularly relevant, but it comes with challenges. In
particular, if the approach to verify correctness lays on the assumption that
the device would be able to execute correctly the output of certain input
instances, we need to be able to answer the following question
\cite{Gheorghiu2018}: ``If a quantum experiment solves a problem which is proven
to be intractable for classical computers, how can one verify the outcome of the
experiment?". Or, as the Aaronson's \$25.00 prize states \cite{AaronsonPrize}:
``If a quantum computer can efficiently solve a problem, can it also efficiently
convince an observer that the solution is correct? More formally, does every
language in the class of quantumly tractable problems (BQP) admit an interactive
proof where the prover is in BQP and the verifier is in the class of classically
tractable problems (BPP)?". Unfortunately, no one has claimed the prize yet.
Still, there are other approaches worth flagging in the field leveraging quantum
provers as trusted measurement devices \cite{Mahadev2018}. 

Related to verification, we can find the concept of verification techniques for quantum benchmarks:
\begin{definition}[Quantum verification technique]
    Any procedure that aims to produce efficiently verifiable results of a quantum experiment.
\end{definition}

In this category, we can find approaches such as the circuit mirroring
\cite{Proctor2021}, or quantum state tomography \cite{PhysRevA.40.2847}, which
can be leveraged to help quantum benchmark designers when trying to measure the
results of their computations. 

We need to clarify that quantum verification and the quantum benchmarking
verification techniques are different from the classic concept of formal
verification of programs. Indeed, there exists a parallel research field
dedicated to the formal verification of quantum programs
\cite{FormalVerificationQuantumPrograms}, and we can easily extend the notion
from their classic counterparts. Note that even though the formal verification
of programs is an interesting field with intensive research for both classical
and quantum cases, historically a testing-based approach that does not rely on
any mathematical formalization for the proof of correctness has been favoured.
For those cases, we introduce the idea of testing quantum programs:
\begin{definition}[Quantum Testing]
\label{Definition.QuantumTesting}
   Process of checking the behavior of a quantum program to confirm it matches
   expected requirements.
\end{definition}

There are some early works in the field of testing \cite{GarcadelaBarrera2021}.
Some of them raise awareness about the challenges of introducing classical
computing techniques for testing such as white- and black-box testing
\cite{8805685}, and hence the need for an enhanced approach for quantum
computers. The importance of quantum testing will become greater in the upcoming
years as we transition to the next eras. 

Finally, we would like to call out that there could be approaches to define
benchmarks that allow the definition of families of circuits or programs, or
they can provide a specification for a subfamily of those. For these cases, we
introduce the definition below:
\begin{definition}[Quantum Benchmark Framework/Family]
    A structure or specification on how to define a family of tests for a benchmark.
\end{definition}
The most well-known example in this category would be the volumetric benchmark framework \cite{Blume_Kohout_2020}.

Also, as mentioned in this article, it is important to highlight that no single
metric can capture the whole behaviour of any classical or quantum computing
device. Hence, to overcome the danger of focusing too much on a single aspect,
the concept of \textit{benchmark suites} can help us:
\begin{definition}[Quantum Benchmark Suite] \label{Definition.QuantumBenchmarkSuite}
   A collection of quantum benchmarks, generally aiming to model different
   aspects of the behavior of a quantum computer. 
\end{definition}

\subsection{Reviewing existing proposals for benchmarking} 
\label{Reviewing}

Recently, there have been attempts to categorize benchmarking practices.
However, the lack of a precise definition of concepts has led to confusion in
the field of quantum benchmarking. There are instances where the concepts of
certification and benchmarking are intertwined within a shared framework
\cite{eisert2020quantum}, or concepts such as metrics and benchmarks are
sometimes used interchangeably \cite{SoKBenchmarkingQuantumComputer}. In Ref.
\cite{SoKBenchmarkingQuantumComputer}, the term \textit{physical benchmarkings}
has been introduced to describe properties of a quantum processor that are more
accurately understood as metrics. Additionally, the concept of
\textit{aggregated benchmarks} has been presented, which combines metrics and
benchmarking techniques. For instance, Circuit Layer Operations per Second
\cite{https://doi.org/10.48550/arxiv.2110.14108} is a metric while Mirroring
Circuits \cite{Proctor2021}) is a technique employed in benchmarks. Other
proposal more aligned with the intent of distinguishing between different terms
have been presented \cite{proctor2024benchmarkingquantumcomputers}. An excellent
example of such studies is the article by Resch and Karpuzcu on the impact of
quantum noise \cite{https://doi.org/10.48550/arxiv.1912.00546}.

On the other hand, in Ref. \cite{https://doi.org/10.48550/arxiv.2110.14108} the
researchers from the IBM T. J. Watson Research Center took a different approach.
They introduced the \textit{benchmarking Pyramid} which classified in three
levels the different benchmarks: Holistic (measuring the systems as a whole),
Subsystem (measuring different aspects of the subcomponents used to assemble the
system, e.g. classical control logic system), and Device (e.g. quantum gates or
qubit registers). Apart from classifying them into these three level groups,
they also distinguished two main factors that can be measured for each level:
Quality (a property that determines the size of quantum circuits that can be
faithfully executed), and Speed (a property related to the number of circuits
that the quantum computer system can execute per unit of time). The
hierarchization process carried out is interesting, and it accounts for a
division of different types of, what we call, \textit{metrics}. Still, during
their work they focused on a procedure to generate a particular test for Quantum
Volume, later defining metrics such as CLOPS on top of it but limiting the scope
of the benchmark. Indeed, there are well-known extensions to the Quantum Volume
to build a family of \textit{volumetric benchmarks} that generalise these types
of tests \cite{Blume_Kohout_2020}. Other remarkable attempts to build thorough
benchmarks can be found in the literature, such as the Application-Oriented
Performance Benchmarks from the Quantum Economic Development Consortium
(QED-C) \cite{quantum_consortium,lubinski2023applicationoriented},
the proposal from Daniel et al.
\cite{Mills2021} where they proposed measuring the performance based on three
types of circuits (deep, shallow, and square classes) with three metrics that
can be computed with classical computers (heavy output generation probability,
cross-entropy difference, and $\ell_1$-Norm Distance), or the QUARK framework
\cite{Finzgar_2022}.

There has been a lack of a standardized framework for assessing the efficacy of
benchmarking within the field. The absence of a common and accepted standard to
define benchmarking techniques and processes for quantum computing has given
rise to a multitude of divergent terminologies, as well as inconsistent
approaches in the field. This work aims to fill this gap by introducing some of
the most prominent techniques, metrics, and benchmarks available as of the
current writing, conducting a comprehensive comparison of these elements using
the framework presented in this paper. Next, we will revisit some relevant
techniques, metrics, and benchmarks, mapping the concepts to the common
framework introduced in this work.

\subsubsection{Techniques}

Let us introduce some relevant techniques that are employed in benchmarks and
metrics, but they are not benchmarks or metrics by themselves.
 \vspace{.3cm}

{\bf Mirror circuits:} The concept of mirror circuits
\cite{Proctor2021,PhysRevLett.129.150502}, inspired by Loschmidt echo, involves
creating an alternative efficient classically simulatable circuit that mirrors a
given quantum circuit $C$. The mirror circuit technique generally consists of implementing a circuit $C$ followed by its inverse $C^{-1}$. There are several variations or specifications of this method. One common approach involves preparing the qubits in randomized Pauli eigenstates, executing the base circuit $C$, applying a randomly selected Pauli layer $Q$, running the reversed circuit $C^{-1}$ adjusted to implement a specific Pauli operation, and finally measuring the qubits in their initial basis. Another variation lies in how $C^{-1}$ is implemented, either by reversing all the gates in $C$, or by applying a different random transpilation of the circuit.

This approach is particularly effective for circuits containing
Clifford and $Z(\theta)$ gates, with $\theta$ being an integer
multiple of $\pi/4$ in $Z(\theta)$ gates, a universal gate set widely used in quantum
processors. Through this method, each circuit in the set
of mirror circuits has a straightforward and computable correct outcome,
allowing the observed probabilities from these circuits to accurately represent
the base circuit's performance.

This technique can be employed in benchmarking to define a metric by generating
a mirror circuit from one or more quantum circuits, probing different aspects of
processor capability. The choice of base circuits, the number of qubits, the
depth, and the structure, influence the benchmarks' focus. In experiments,
benchmarks were created from both disordered and highly structured periodic base
circuits using native gates, respecting processor connectivity and implementing
elements of the Clifford group.

 \vspace{.3cm}
    
{\bf Classical shadow tomography:} Classical shadow tomography addresses
challenges in efficient information extraction from quantum states
\cite{doi:10.1137/18M120275X,article12,zhou2023efficient}. Traditional methods
demand an exponential number of measurements for a complete description of
$N$-qubit states, which is impractical as discussed in
Section~\ref{subsection:nuances of QC}. However, classical shadow tomography
provides an efficient alternative for extracting information with a limited
number of measurements when detailed knowledge of the entire density matrix is
unnecessary. This technique uses randomized measurements
\cite{zhou2023efficient}, employing independent random unitary operations, $U$,
selected from an ensemble. Before measurement, this operation is applied,
yielding a classical snapshot $\sigma_{U,b}=U^\dagger|b\rangle\langle b|U$ with
a probability $p_{U,b}$. The average snapshot $\sigma$ relates to the original
density matrix $\rho$ through a quantum channel $M$ as $\sigma = \sum_b
\mathbb{E}_{\mathcal{E}_u}[\sigma_{U,b}p_{U,b}] = M(\rho)$. Physical
observables, denoted as $\langle O\rangle$, can be expressed as $\langle
O\rangle=\text{tr}\left(OM^{-1}(\sigma)\right)$. Consequently, efficient
classical computation of $M^{-1}$ and a low sampling complexity are key criteria
for effective classical shadow tomography schemes. Different schemes correspond
to diverse choices of the ensemble $\mathcal{E}_U$. For instance, utilizing Haar
random unitaries and tensor products of single-qubit Haar random unitaries, one
can prove that while the shadow norm $\|O\|^2_{\text{sh}}$ scales as $2^N$ for
general $N$-qubit systems, it scales as $3^k$ for Pauli operators with a size of
$k$, showing a better performance for few-body operators.

 \vspace{.3cm}

{\bf Efficient classically simulatable quantum circuits:} Several families of
quantum circuits can be efficiently simulated in classical computers.
Particularly relevant are Clifford circuits, a specific class of quantum
circuits composed entirely of Clifford gates. These gates include the Hadamard
gate ($H$), the Phase gate ($S = \sqrt Z$), and the Controlled-NOT gate
($\text{CNOT}$). The efficient simulatability of these circuits is ensured by
Gottesman-Knill theorem, which states that any quantum circuit comprising an
initial state in the computational basis, Clifford gates, and measurement in the
computational basis, can be efficiently simulated in a classical computer
\cite{10.5555/863284,nielsen00}.

\vspace{.3cm}

 {\bf Randomized Benchmarking (RB):} Randomized Benchmarking (RB) is a technique used for characterizing the error rate of quantum processors. This technique involves applying a sequence of random gates, which can be sampled uniformly at random from the Haar distribution \cite{Emerson_2005} or uniformly at random from any unitary 2-design \cite{Dankert_2009}, such as the Clifford group. RB has been employed as a technique for measuring errors, such as the gate fidelity of Clifford gates \cite{PhysRevLett.106.180504}, the coherent or incoherent nature of the noise affecting Clifford gates \cite{Wallman_2015}, or the noise of a particular gate \cite{Magesan_2012}, among others. This technique is robust against state preparation and measurement (SPAM) noise and is considered scalable. Although the standard randomized benchmarking (RB) protocol employing this technique is theoretically scalable, in practice it becomes challenging to implement for more than 5 or 6 qubits due to the depth of the required Clifford circuits. To overcome this limitation, recent variants of the protocol have been developed that aim to minimize circuit depth, making the method viable for systems with up to 20 qubits \cite{Hines2024, PhysRevLett.129.150502, polloreno2023}. For a detailed review of the different types of protocols based on this technique, see \cite{hashim2024}.

\subsubsection{Metrics}
\label{QMetrics}
Let us introduce some of the main metrics that have been proposed to measure
properties related to the performance of a quantum computer. We will categorize
these metrics into two types: Static and Non-static metrics. Static metrics refer to internal features of the quantum processor that define
specific performance-related properties. Once the processor is manufactured,
these metrics cannot be altered. 

For each metric, we indicate which properties from \ref{enum:good_metric} it satisfies. We do not attempt to quantify the degree of satisfaction of each property, as such grading is difficult to define and not essential for the purposes of this work. Accordingly, our classification records only whether a property is satisfied or not. However, when a property is not satisfied, this may occur to varying degrees and for different reasons, including platform heterogeneity, strong dependence on the experimental context, unfavorable scalability, limited ability of the metric to reliably discriminate device performance, or ambiguities in the metric definition itself, among other factors. Next, we introduce the main static metrics:

 \vspace{.3cm}
 
 {\bf Number of qubits:} One of the most commonly used metrics in quantum
 performance evaluation is the number of (working connected) qubits, which
 directly impacts the quantum information processing capacity
 \cite{https://doi.org/10.48550/arxiv.2110.14108}. As the number of qubits
 increases, the potential of the quantum computer to address larger and more
 complex problems grows exponentially. We emphasize the importance of
 considering not only the quantity but also the quality of qubits, which is
 especially relevant in the NISQ era. Indeed, scaling quantum computers to a
 large number of qubits is a challenging engineering task that involves
 addressing various technical issues depending on the characteristics of the
 physical platform in use.When the metric is defined as the number of qubits in a quantum computer that form a connected graph, it trivially satisfies all the properties required of a good metric.
 
 \vspace{.3cm}

{\bf Qubits connectivity:} Qubit connectivity is an important metric to consider
because it determines the direct interactions between the qubits
\cite{Holmes_2020}. Quantum algorithms require the interaction of multiple
qubits through quantum gates. When qubits are not directly connected additional
quantum gates could be necessary to facilitate their interaction increasing the
runtime of the algorithm, circuit depth, and the risk of errors in the NISQ era.
Therefore, qubit connectivity directly affects the performance of the quantum
computer. Currently, physical architectures provide varying qubit connectivity
options, such as nearest-neighbor connectivity
\cite{Arute2019,ZHU2022240,PhysRevLett.127.180501} or fully connected
connectivity \cite{QCion,Zhang_2017}. One could provide the maximal, minimal and
average connectivity, for instance. Other method of mapping the qubit
connectivity into a numerical representation involves constructing a $n\times n$
matrix of normalized coupling terms for a $n$-qubit processor and compute its
norm. This approach has as an advantage that it takes into account also the {\it
quality} of the couplings but the norm complicates comparing processors with
different sizes. When considering the average connectivity, the metric is clearly practical and repeatable, as it depends only on fixed quantities that do not vary over time. However, it is not reliable, since performance is influenced not only by the number of connections but also by how those connections are arranged and by the quality of the connections. As a result, a quantum computer with fewer connections may allow for shallower circuits than another system with higher average connectivity. For the same reason, the metric is not linear. Finally, it is consistent, as its definition does not depend on the underlying hardware architecture.
\vspace{.3cm}

On the other hand, non-static metrics also characterize a performance-related
property of the quantum processor. Unlike static metrics, after manufacturing
the quantum processor, these metrics can be altered according to their
application. Next, we present the main non-static metrics:

\vspace{.3cm}
{\bf Gate fidelity:} The gate fidelity refers to the accuracy of quantum
operations. There is no unique definition of fidelity for a quantum operation; however, most definitions are linearly related and share similar properties. Therefore, throughout this text, we refer to fidelity in a general sense. In practical applications, however, it is important that users specify the particular fidelity measure being used. The gate fidelity focuses on how well a
quantum gate performs the intended operation compared to the ideal outcome.
When
a quantum gate is applied, there is a predetermined ideal output, but due to
various factors like noise and imperfections in the quantum processor, the
actual observed output might deviate from the expected one. High gate fidelity
is essential for the reliability and precision of computations. If the gate
fidelity is low, it indicates that the quantum gate is introducing errors,
reducing the accuracy of quantum algorithms. 
However, although the concept is
clear, its practical application in a quantum processor presents some complex
problems. In quantum platforms where entangling gates are natively implemented by two-qubit gates, such as in superconducting circuits or photonic quantum computing, it is common practice to characterize the performance of quantum hardware using metrics associated with two-qubit gates. In such cases, it is natural to evaluate the average, maximum, or minimum fidelity of these gates. However, this logic does not directly apply to other platforms, such as trapped-ion or cold-atom systems, where native multi-qubit gates can be implemented. In these architectures, a single operation can generate collective entanglement among three or more qubits without the need to decompose the operation into a sequence of two-qubit gates. This capability fundamentally change the structure of quantum computation and the efficiency with which certain algorithms can be executed. While comparing individual two-qubit gate fidelities across platforms can be valid when the same operations are implemented, relying solely on such metrics to assess the overall performance of a quantum processor can be misleading. This is particularly true when native gate sets differ significantly, as is the case between platforms supporting multi-qubit gates and those limited to two-qubit gates. In this case, moreover, the fidelity is substantially more costly to
calculate. It is also important to note that gate fidelity is not a static property of a quantum hardware. It can vary over time due to coherence times of qubits that are highly sensitive to environmental. Additionally, the fidelity depends on implementation details such as pulse shaping and calibration routines, which can be optimized without manufacture a new quantum hardware. When considering the average gate fidelity of one- and two-qubit gates, which is the metric most commonly used in practice, the metric is practical, as gate fidelities can be efficiently estimated for systems involving at most two qubits. It is also repeatable when evaluated under the same or comparable hardware conditions. However, the metric is not reliable, since it is an average quantity: a higher average gate fidelity does not necessarily imply higher overall computational quality, as a device may achieve high fidelity only for a restricted class of circuits. The metric would be consistent if the average were computed over a common set of generic one- and two-qubit gates. In practice, however, experimental evaluations typically average over native gates, which differ across platforms and architectures. As a result, average gate fidelities cannot be directly compared across different quantum computers. Finally, we do not assess linearity for this metric, as there is no clear notion of how to linearly quantify the overall ``quality'' of a quantum computer, in contrast to metrics such as execution speed.

\vspace{.3cm}  
{\bf Readout fidelity:} The readout fidelity is a metric that quantifies the accuracy with which a quantum processor measures the final state of qubits. It reflects the probability that the measurement outcome corresponds to the quantum state, typically in the computational basis. High readout fidelity is essential for reliable data collection and accurate measurement of qubit states, especially in NISQ devices, where measurement errors can significantly impact overall performance. Importantly, several techniques can be employed to improve readout fidelity without the need to manufacture a new quantum processor \cite{PRXQuantum.5.040326,PhysRevApplied.23.054057,Seif_2018}. This metric is practical because it can be efficiently estimated by repeatedly preparing $\ket{0}$ and $\ket{1}$ and measuring the corresponding assignment errors. It is repeatable since repeating the procedure under similar conditions yields consistent values within statistical fluctuations. It is reliable when the goal is to evaluate the quality of the measurement stage. However, it is not a fully reliable indicator of the overall computational performance of a quantum processor, because a device may exhibit excellent readout fidelity while still suffering from poor gate fidelities, limited coherence times, or other features that affect processor performance. It is consistent because it is defined in the same way across different hardware architectures.

\vspace{.3cm}  
{\bf Decoherence times:} The decoherence times refer to the period during which
the qubit can maintain its quantum coherence. The interaction of qubits with the
environment results in the loss of quantum superposition and the decay from the
excited state to the ground state, characterized by the times $T_2$ and $T_1$
respectively. Longer decoherence times are preferable as they might permit
deeper circuits to be run, reducing the probability of errors and improving
computation accuracy. Additionally, techniques such as dynamical decoupling \cite{Viola1998} can be used to improve coherence times by mitigating specific sources of noise applying control pulses to qubits. This metric is practical because $T_1$ and $T_2$ can be efficiently estimated using standard procedures with low experimental overhead. It is repeatable since repeating the same protocols under similar conditions yields comparable values within statistical fluctuations. It is reliable for characterizing intrinsic coherence limitations, as shorter $T_1$ and $T_2$ times generally imply faster loss of quantum information and reduced circuit depth capability. However, it is not a complete predictor of quantum processor performance during the execution of a quantum circuit, because gate errors, crosstalk, control imperfections, and measurement errors may dominate even when coherence times are long. It is not a linear metric, since increasing $T_1$ or $T_2$ does not translate proportionally into higher quantum circuit accuracy or overall quantum processor performance. The metric is consistent because $T_1$ and $T_2$ have a clear physical meaning and are defined uniformly across platforms.

\vspace{.3cm}
{\bf Gate speeds:} Gate speed is a metric that measures how rapidly a quantum
computer can execute a given quantum gate \cite{van_der_Schoot_2023}.
Considering decoherence times, this metric allows us to determine the number of
gates that can be performed before the qubits lose coherence. By increasing gate speeds, a greater number of gates can be executed within the same time interval,
leading to more applied gates during the lifetime of the qubits. Here, the
argument explained for coherence times can be inverted, since fast manipulable
systems could also be susceptible to decoherence. The problems of this metric
are similar or worse to gate fidelity metric. It is important to highlight that gate speed is not a fixed metric, as it can be optimized through pulse engineering techniques \cite{Werninghaus2021}. Moreover, quantum computing platforms, the strength of the interaction between qubits can be dynamically tuned via external control parameters \cite{Yan2018}. Therefore, both pulse engineering techniques and dynamic tuning of qubit interactions provide control over quantum gate speeds. This metric is practical, as it only requires executing the quantum gate and measuring its execution time. It is repeatable when the same circuits are run multiple times under comparable experimental conditions. However, it is not reliable, since gate speed is typically evaluated over the native gate set of a quantum computer. Consequently, a device that performs better under this metric does not necessarily execute arbitrary circuits faster than another, as performance depends on how circuits are decomposed into native gates. Moreover, overall performance also depends on qubit coherence times, and faster gates do not necessarily imply better computational performance. For the same reason, the metric is not consistent across different platforms. Finally, the metric is linear: if one system achieves a value that is a factor $x$ smaller than another, it will execute the same circuits in $x$ times less time.

\vspace{.3cm}
 
{\bf Quantum Volume (QV):} This is a metric that quantifies the largest square
(equal in width and layers) quantum circuit that a quantum processor can
successfully run \cite{PhysRevA.100.032328}. Each layer consists of random
permutation among qubits followed by pseudo-random 2-qubit gates. After running
the random quantum circuit, the measured results should be above a certain
threshold to be considered successful. Consequently, higher-quality quantum
gates and qubits allegedly result in a larger QV. More specifically, ${\rm
QV}=2^D$, where $D$ is the largest square successfully run. For example, a
quantum volume 256 means that a quantum processor can use a subset of 8 qubits
in the processor to successfully run a circuit with 8 sequential layers. A
higher QV indicates a greater capability of the quantum processor to tackle more
complex tasks, demonstrating advancements in the quality of qubits and error
mitigation techniques. However, one limitation of QV is that it requires knowing
the probabilities of all possible outcomes, which implies a large number of
measurements that scale exponentially with the number of qubits. Another
limitation is that the QV protocol exclusively considers square circuits.
Therefore, QV does not offer a reliable prediction of the performance of
circuits that have high depth and low width, or circuits with low depth and high
width. Due to the strong limitations of this metric, IBM proposed the
alternative Error Per Layered Gate (EPLG)
\cite{mckay2023benchmarkingquantumprocessorperformance}. This metric makes use
of (Clifford) randomized benchmarking to compute
the layer fidelity LF, and then $\text{EPLG} = 1- \text{LF}^{1/n_{2Q}}$, where
$n_{2Q}$ is the number of two-qubit gates, tipically taking the value $N-1$ for
a linear chain of qubits. The exponent is used to convert the aggregate process fidelity of a layer into a “per-gate” error rate by taking the geometric mean. This gives rise to the term Error Per Layered Gate. As discussed above, the Quantum Volume (QV) metric is not practical, as it requires an exponential number of operations to be evaluated. It is repeatable when the same circuits are executed a sufficient number of times under comparable hardware conditions. However, it is not reliable, since it does not fully characterize the general computational performance of a quantum processor, as previously discussed. The metric is not linear, as it scales exponentially by construction. Finally, it is consistent, since the task of implementing random unitaries can be realized across different platforms.

\vspace{.3cm}
 
{\bf Q-Score:} Q-score is an application-focused metric which measures the
maximum number of variables that a quantum processor can optimize on a standard
combinatorial approximate optimization problem (for instance, a Max-Cut Problem)
\cite{9459509}. Therefore, Q-Score is remarkably an application-oriented metric,
since it focuses on solving a practical task. Another advantage is the
versatility of the Q-score, as it can be executed and computed on any quantum
processor, although in the NISQ era highly connected processors, such as the
ones in trapped-ion platforms, are favored \cite{doi:10.1126/sciadv.aau0823}. On
the other hand, this versatility does facilitate the comparison of devices
between different computational paradigms, such as quantum annealers and
gate-based \cite{vanderschoot2023qscore}. This metric is practical, as the comparison relies on the optimized value obtained using a heuristic method. It is also repeatable when the same compilation rules and circuits are used, provided that a sufficiently large number of measurements is performed. The metric is reliable for comparing how effectively a quantum computer optimizes specific classes of graphs using the QAOA technique; however, it is not reliable for more general graph families or for generic optimization problems addressed with this approach. The metric is not linear, since an improvement by a factor $x$ in the metric value does not imply the ability to solve the problem $x$ times better. Finally, the metric is consistent, as it does not rely on device-specific evaluation rules and can be applied across different architectures and platforms.

\vspace{.3cm}

{\bf  Circuit Layer Operations Per Second (CLOPS):} This metric, proposed as a
natural extension of Floating Point Operations per Second (FLOPS) in
supercomputers, quantifies the number of QV-type layers that the quantum
processor can run per second \cite{wack2021quality}. This metric was proposed by
IBM to measure the performance of a processor in terms of speed, complementing
QV as a metric, which was focused on measuring quality. Remarkably, CLOPS
include also the classical processing and control times in the calculation.
Therefore, a larger CLOPS value indicates a higher speed of the quantum
processor in executing layers of the same type used in the QV. Therefore, this
metric suffers from limitations similar to those of QV regarding scalability,
making it {\color{blue} not practical}. Additionally, the metric does not naturally extend to processors that support native multi-qubit gates, such as trapped-ion hardware, as it assumes a circuit structure composed exclusively of one and two-qubit operations. Consequently, its application to architectures with fundamentally different native gate may lead to inconsistencies in performance comparisons {\color{blue} making it not consistent}. Finally, as CLOPS include the time corresponding to
control, one can improve it by faster classical electronics, without actually
improving the processor, so reliability {\color{blue}and linearity} might also be in question. Due to these
limitations, IBM proposed an update of this metric called CLOPS$_h$
\cite{www.ibm.com}
which updates the definition of {\it layer} and now the possible two-qubit gates
are no longer arbitrary, but they are only the gates which are native in the
processor (i.e. between qubits which are directly connected). The reason is
other gates must be decomposed into native ones, jeopardizing the definition of
layer as one composed by parallel gates. Although this update is reasonable, it
does not fix any of the problems associated to CLOPS.

 \vspace{.3cm}
 
{\bf Algorithmic Qubits (AQ):} Algorithmic Qubits (AQ=$N$) is an
application-oriented metric that quantifies the largest quantum circuit of width
$N$ and depth $N^2$ that a quantum processor can successfully run
\cite{chen2023benchmarking}. In contrast to quantum volume, the AQ metric does
not concentrate on random quantum circuits but instead on six algorithmic
classes, including Hamiltonian simulation, phase estimation, quantum Fourier
transform, amplitude estimation, variational quantum eigensolver (VQE)
simulation, and Monte Carlo sampling. The success of the quantum circuit is
measured by evaluating the fidelity, which involves comparing the ideal output
probability distribution to the measured output distribution on the quantum
computer. It is considered successful only if it surpasses a certain threshold.
Consequently, one must know a priori the outcome of the running algorithms,
which is problematic for sufficiently large processors which cannot be
classically simulated. Additionally, as the success is measured by the fidelity,
it also shows scalability problems making the metric not practical. The metric is repeatable for the same reasons as the Q-Score. However, it is not reliable for general problems, including the six instances considered here, due to limitations in achievable circuit depth. The metric is not linear, as the required circuit depth scales quadratically. Finally, the metric is consistent, since it can be applied across different computational paradigms and hardware platforms.

 \vspace{.3cm}

{\bf Cross-entropy difference and cross-entropy benchmarking (XEB):} This metric
has been used to quantify the correspondence between the experimental and the ideal output
probability distributions of a pseudo-random quantum circuit \cite{Qsupremacy}, and it has been used to calculate the XEB fidelity \cite{Arute2019}. The cross-entropy
difference $\Delta H (p_\mathrm{A})$ is defined as 
\begin{equation}
    \Delta H(p_\mathrm{exp})=H(p_\mathrm{cl},p) - H(p_\mathrm{exp},p),
\end{equation}
where $H(p_\mathrm{1},p_\mathrm{2})=-\sum^N_{i=1} p_{1}(x_i)\log{p_2(x_i)}$ is the cross-entropy
between the probabilities distributions $p_1$ and $p_2$, and $N$ represents
the total number of possible outputs. Additionally, the experimental, ideal, and
uniform probability distributions are denoted by $p_\mathrm{exp}$, $p$, and $p_\mathrm{cl}$,
respectively. In practice, we will have an experimental implementation of the
Haar random unitary $U$ as $A_\mathrm{exp(U)}$ with an associated probability
distribution $p_\mathrm{exp}$ and sample $S_\mathrm{exp} = \{x_1^\mathrm{exp}, \dots x_m^\mathrm{exp}\}$. In
this case, the (ideal) experimental cross-entropy difference is defined as
\begin{equation}
    \alpha \equiv \mathbb{E}_\mathrm{U}(\Delta H(p_\mathrm{exp})),
\end{equation}
where $\mathbb{E}_\mathrm{U}$ is the expected value over the Haar random unitaries
distribution. This $\alpha$ is now estimated from an experimental sample of
bitstrings $S_\mathrm{exp}$ obtained by measuring the output of $A_\mathrm{exp(U)}$ after $m$
realizations of the circuit, as
\begin{equation}
    \alpha \simeq \tilde{\alpha} = H(p_\mathrm{cl},p_\mathrm{U}) - \frac{1}{m}\sum_{j=1}^m \log
    \frac{1}{p_U\left(x_j^\mathrm{exp}\right)},
\end{equation}
with an error of $k/\sqrt{m}$ and $k\simeq 1$. Here, $p_\mathrm{U}\left(x_j^\mathrm{exp}\right)$
is the ideal probability of the $x_j^\mathrm{exp}$ bitstring computed classically. With $m \sim 10^3 - 10^6$ number of samples, a sufficiently accurate approximation can be achieved. The (ideal) experimental cross-entropy difference yields a value ranging from
$0$ to $1$, which approximates the average gate fidelity of the circuit. When
the experimental output probability distribution aligns perfectly with a uniform
distribution, it represents the worst-case scenario, obtaining an experimental cross entropy
difference equal to zero. On the other hand, when the experimental
output distribution perfectly matches the ideal distribution, the
experimental cross-entropy difference is one. It is worth emphasizing that
experimentally generating random circuits poses a significant challenge.
Moreover, obtaining the ideal probability distribution requires classical
computational resources that scale exponentially with the number of qubits,
making this metric not practical. To discuss whether this metric is repeatable
and reliable, it is important to distinguish between the theoretical method used
to calculate $\alpha$ and the experimental implementation used to compute
$\tilde{\alpha}$. In the theoretical method, the metric satisfies the repeatable
and reliable characteristics. In the experimental implementation, there are two
options: one can take a fixed finite set of Haar random unitaries, using
always this set, or generate a new set of random unitaries every time. The first case
satisfies both properties, but the second does not.  However, this is more
associated with the benchmarking methodology than the metric itself, so we will
analyze the metric according to the theoretical method. Finally, the metric is consistent, as its definition does not depend on any specific platform or hardware architecture.

 \vspace{.3cm}
 
{\bf Hellinger distance:} The Hellinger distance is a metric used to measure the
similarity between two probability distributions. It has been used to quantify
the capability of a quantum computer to accurately replicate the outcomes of a
specified quantum circuit \cite{e24020244}. The Hellinger distance between the
ideal probability distribution $P^\mathrm{ideal}$ and the experimental output
distribution $P^\mathrm{exp}$ is defined by 
\begin{equation}
    d(P^\mathrm{ideal},P^\mathrm{exp})=\sqrt{1-BC(P^\mathrm{ideal},P^\mathrm{exp})},
\end{equation}
with the Bhattacharyya coefficient $BC(P^\mathrm{ideal}, P^\mathrm{exp}) \in [0, 1]$ defined
as:
\begin{equation}
 BC(P^\mathrm{ideal},P^\mathrm{exp})= \sum_{i=1}^N \sqrt{p^\mathrm{ideal}_i p^\mathrm{exp}_i}. 
\end{equation}
The ideal and experimental probabilities to observe the $i$-th outcome are
denoted by $p_i^\mathrm{ideal}$ and $p_i^\mathrm{exp}$, respectively. The number of possible
outcomes is given by $N=2^n$, where $n$ is the number of qubits. The Hellinger
distance converges to $0$ for identical distributions and $1$ in cases, where
there is no overlap between distributions. Since computing the ideal output distribution requires classical simulation of the quantum circuit, the metric is not practical. Nevertheless, under its theoretical formulation it is repeatable and reliable, and it is consistent because its definition is independent of the underlying hardware or platform. Furthermore, it is important to note that the Hellinger distance, like other classical metrics such $\ell_1$-Norm Distance, depends on the measurement basis and therefore may fail to capture features such as phase errors.

\vspace{.3cm}
 
{\bf Heavy Output Generation (HOG):} The (HOG) metric is a measure between two probability distributions, and in the context of quantum computing, it has been used to quantify the
success of a pseudo-random quantum circuit executed on a quantum computer
\cite{10.5555/3135595.3135617}. If we consider a pseudo-random circuit $U$ taken
from a uniform Haar distribution, the ideal and experimental output probability
of each bitstring after executing $U$ is denoted by $P^\mathrm{ideal}$ and
$P^\mathrm{exp}$. A bitstring $z$ is classified as {\it heavy} in $U$, if its
probability $P^\mathrm{exp}(z)$ is greater than the median of the distribution
$P^\mathrm{ideal}$. The probability that a sample drawn from a distribution
$P^\mathrm{exp}$ is heavy, named as the {\it heavy output generation
probability} of $P^\mathrm{exp}$, is given by
\begin{equation}
    HOG(P^\mathrm{exp},P^\mathrm{ideal}) = \sum_{i=1}^{N} p^\mathrm{exp}_i \delta_i,     
\end{equation}
where $N = 2^n$, $n$ is the number of qubits, and $\delta_i = 1$ if $i$ is heavy
in $U$, and $0$ otherwise. If $P^\mathrm{exp}$ approximates the uniform
distribution, $\mathrm{HOG}(P^\mathrm{exp},P^\mathrm{ideal}) \approx 1/2$; and
if $P^\mathrm{exp}$ is close to $P^\mathrm{ideal}$,
$\mathrm{HOG}(P^\mathrm{exp},P^\mathrm{ideal}) \approx (1+\log(2))/2$. The
$\mathrm{HOG}$ metric has been employed to develop a test that assesses the
performance of a quantum computer as {\it good} if it exceeds $2/3$. One
limitation of $\mathrm{HOG}$ is that it requires calculating the ideal
probability classically, which makes this metric not practical. To assess whether this metric is repeatable, reliable, and consistent, we follow the same criteria used for the cross-entropy difference and XEB metrics. Under these criteria, the metric satisfies all these properties.
\vspace{.3cm}
 
{\bf $\ell_1$-Norm Distance:} The $\ell_1$-Norm Distance is a metric that
quantifies the total difference between the output distribution of the quantum
circuit produced by a quantum computer, denoted by $P^\mathrm{exp}$, and the
probability output distribution of the ideal quantum circuit generated by a
classical computer, denoted by $P^\mathrm{exp}$ \cite{Mills_2021}. The $\ell_1$-
norm distance reads as 
\begin{equation}
    \ell_1 (P^\mathrm{ideal},P^\mathrm{exp})=\sum_{i=1}^{N} |p^\mathrm{ideal}_i-p^\mathrm{exp}_i|,
\end{equation}
where $p_i^\mathrm{ideal}$ and $p_i^\mathrm{exp}$ are the ideal and experimental
probabilities to observe the $i$-th outcome and $N=2^n$ is the number of
possible outcomes, with $n$ the number of qubits. The classical resources required to obtain the ideal probability distribution scale exponentially with the number of qubits, rendering this metric not practical. For the same reasons as discussed for the Hellinger distance, this metric satisfies the same set of properties.

\vspace{.3cm}
 
{\bf Collision volume:} This metric has been used to quantify the noise of a
quantum computer using Haar random unitaries \cite{mari2023}. Collision volume
involves executing a random circuit $U$, obtaining $m$ samples, and counting how
many bitstrings have been sampled more than one time, what it is called {\it
collision}. It can be proven that in an ideal scenario in which the quantum
computer is not noisy, the number of collisions doubles the case in which the
computer is completely noisy. Then, the collision volume is defined as
\begin{equation}
    \hat{\Delta } \equiv 
    \frac{\hat{R} - N + N(1-e^{-m/N})}{N^2/(N+m) - Ne^{-m/N}},
\end{equation}
where $m$ is the number of samples, $N = 2^n$ with $n$ the number of qubits, and
$\hat{R}$ is the observed number of collisions. The expected value of
$\hat{\Delta}$ is equal to $1$ when sampling from a typical random quantum
state, and $0$ when sampling from a uniform distribution. Using this metric, the
authors propose a test called collision volume test, which indicates that the
computer has {\it good} accuracy if $\hat{\Delta} \geq 1/2$. It is important to
emphasize that the collision volume metric does not require a classical
computer, but it requires around $2^{n/2+5}$ samples to work, which makes it not
practical. To determine if this metric is repeatable, reliable and consistent, we follow the same criteria used in the cross-entropy difference and XEB metric, which means that it satisfy these quality attributes.

\subsubsection{Benchmarks in NISQ Era}
\label{Bench}
Let us now review some relevant benchmarks that have been employed to assess the
performance of a quantum processor.

\vspace{.3cm}

{\bf QPack:} The QPack benchmark utilizes Quantum Approximate Optimization
Algorithms (QAOA) to evaluate the effectiveness of quantum computers in solving
problems such as MaxCut, dominating set, and the traveling salesman problem
\cite{QPACK2021}. This benchmark assesses various aspects of a quantum computer,
including runtime, output accuracy, and scalability. The performance of the
quantum computer is assessed by contrasting the solution obtained through QAOA
execution on the quantum computer with that obtained using a classical exact
algorithm. Consequently, the size of the problem that can be handle is
constrained by the computational capacity of the classical computer. The QPack
benchmark has been used to compare different quantum computing platforms, such
as superconducting and trapped ion devices \cite{QPACK_}.

\vspace{.3cm}

{\bf  Quantum LINPACK:} Inspired by the classical LINPACK benchmark, the Quantum
LINPACK benchmark assesses the performance of a quantum computer based on its
ability to solve linear equations system using pseudo-dense random matrices.
\cite{QuantumLINPACK}. The creators of this benchmark introduce an input model
known as the RAndom Circuit Block-Encoded Matrix (RACBEM), which serves as an
appropriate generalization of a dense random matrix in the context of quantum
computing. The success of the benchmark is measured by comparing the output
vector to the ideal solution, incurring an exponential classical cost.

\vspace{.3cm}

{\bf QASMBench:} The QASMBench is a benchmark that includes a set of quantum
circuits and metrics for characterizing circuit properties both before and after
execution on a quantum processor \cite{li2022qasmbench}. The set of quantum
circuits are frequently used in various fields, such as chemistry, simulation,
linear algebra, search algorithms, optimization, arithmetic, machine learning,
fault tolerance, cryptography, and more \cite{QuantumAlgorithms}. QASMBench is
divided into three categories: small-scale, medium-scale, and large-scale, each
defined by the number of qubits used. As an illustration, in the small-scale
scenario, the performance of quantum processes is assessed by executing quantum
algorithms using 1-5 qubits such as Quantum Walk, Linear Equation, Search, and
optimization, among others.  In the medium-scale and large-scale scenarios, the
performance of quantum processes is evaluated by running quantum algorithms
using 6-15 and more than 15 qubits, respectively. The authors propose a set of
circuit evaluation metrics: Circuit Width, Circuit Depth, Gate Density,
Retention lifespan, Measurement Density, and Entanglement Variance, which are
indicators that allow us to quantify the performance of a certain quantum
circuit executed on a quantum processor. QASMBench has been assessed on IBM-Q
machines, and the effectiveness of the quantum circuits is determined by
computing the Hellinger distance. 

\vspace{.3cm}

{\bf SupermarQ Suite:} SupermarQ Suite is a benchmark that includes a set of
applications that assess the performance of a quantum computer based on its
ability to generate entanglement between qubits through the GHZ algorithm,
demonstrate the quantum nature of the system through the Mermin-Bell test, solve
combinatorial optimization problems through the quantum approximate optimization
algorithm, find the ground state energy of the one-Dimensional Transverse Field
Ising Model (1D TFIM) using the Variational Quantum Eigensolver algorithm (VQE)
and simulate the time evolution of the 1D TFIM Hamiltonian using a quantum
circuit via Trotterization for a specific number of time steps \cite{9773202}.
The performance of each application is assessed through different scoring
function, comparing experimental outcomes with predefined ideal results. As
these ideal results are known in advance, the scalability of this benchmark is
ensured. The SupermarQ Suite has been employed to assess the performance of both
superconducting and trapped ion devices.

\vspace{.3cm}
{\bf QED-C Benchmark:} \label{part:qed-c} The Quantum Economic Development Consortium (QED-C)
introduced the QED-C benchmark to assess the efficiency of quantum computers
\cite{lubinski2023applicationoriented}. This benchmark comprises various
algorithms, including  Deutsch-Josza, Grover's Search, VQE, Phase Estimation,
Bernstein-Vazirani, Quantum Fourier Transform, and other algorithms. Unlike the
Quantum Volume metric, which only uses square circuits, the QED-C benchmark
incorporates circuits that vary in both depth and width. It includes circuits
with high depth and low width, as well as circuits with low depth and high
width. The QED-C benchmark evaluates computer performance by assessing fidelity,
which involves comparing the ideal probability distribution with the
experimental probability distribution. Therefore, this approach is limited by
its requirement for exponentially costly classical computations to evaluate the
performance of quantum computers.

\vspace{.3cm}

{\bf Chemistry simulation:} The Chemistry Simulation benchmark is designed to
assess the performance of quantum computers by the electronic structure
calculation of molecules using variational methods \cite{QuantumChemistry}. The
variational quantum eigensolver (VQE) algorithm has been used to calculate
electronic structures using various quantum computing architectures both
photonic \cite{VQE-Photonic}, superconducting circuits \cite{PhysRevX.6.031007}
and trapped ions \cite{PhysRevX.8.031022}. This benchmark assesses the
performance of quantum computers by using ground state energy as a metric,
comparing results obtained on classical computers with those derived using the
VQE algorithm on quantum computers.

\vspace{.3cm}

We have introduced the main proposals of metrics and benchmarks in the
literature. Table \ref{T1} analyzes metrics, evaluating which quality attributes
they fulfill according to the framework introduced in Sec. \ref{Classical
Benchmarking}. We emphasize that we evaluate the quality attributes fulfilled by
metrics in the NISQ era; Therefore, these evaluations are not valid for the
post-NISQ era. We hope that as quantum processors evolve, benchmarks will adapt
to each era.Table~\ref{T1} provides an assessment of quantum computing metrics that have been proposed to measure performance-related properties of a quantum computer. A check mark indicates that the metric satisfies a given quality attribute. A cross indicates that the attribute is not satisfied in general; importantly, cross marks should be interpreted with nuance. In many cases, this may arise from limitations imposed by platform heterogeneity, strong dependence on experimental conditions, unfavorable scalability, an insufficient ability of the metric to reliably discriminate device performance, or ambiguities in the metric definition itself, among other factors. A dash mark indicates that the corresponding attribute cannot be assessed unambiguously for the metric as commonly defined, either because it is not applicable or because its evaluation depends on additional assumptions.

\begin{figure}
    \centering
    \includegraphics[width=\linewidth]{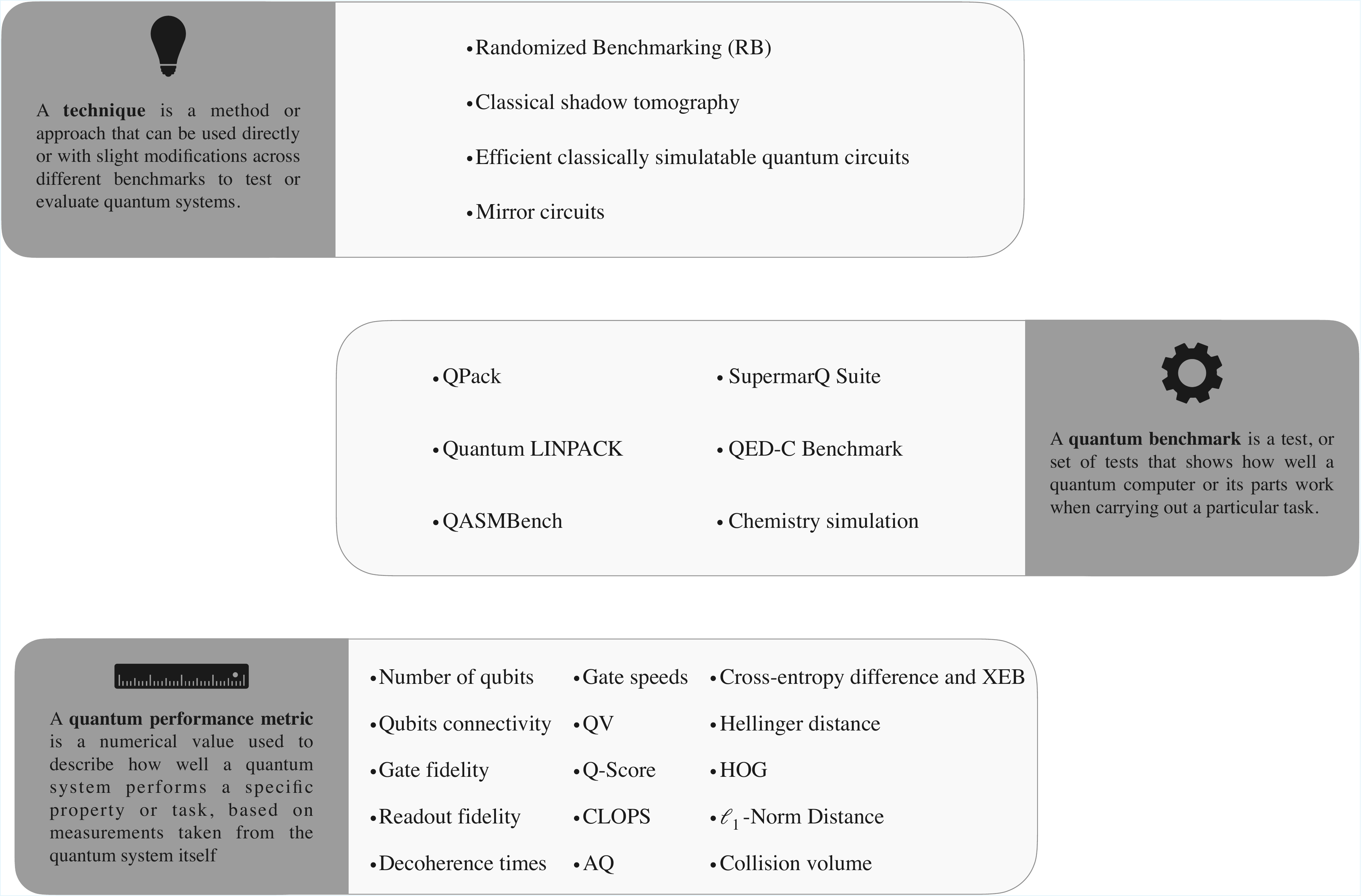}
    \caption{Overview of the terminology used in this work. We distinguish between techniques, quantum benchmarks, and quantum performance metrics. Representative examples of each category are provided.}
    \label{fig:classification}
\end{figure}

\vspace{.3cm}

\begin{table}
\caption{Summary of quantum computing metrics in NISQ era. In this table, we
show some of the main metrics that have been proposed to measure properties
related to the performance of a quantum computer, as presented in subsection
\ref{QMetrics}. We assess whether these metrics satisfy the quality attribute proposed
in the literature for an ideal performance metric, as discussed in subsection
\ref{GoodMetric}, although some of them are borderline. The check mark indicates
that the metric satisfies the attributes; otherwise, we use the cross mark. A cross indicates that the corresponding attribute is not strictly satisfied. This may arise from limitations imposed by platform heterogeneity, strong dependence on experimental context, unfavorable scalability, an insufficient ability of the metric to reliably discriminate device performance, or ambiguities in the metric definition itself, among other factors. The dash mark indicates that it is complicated to define if the metric satisfies the criterion because we cannot characterize it formally.
Performance definitions: {\it Scalability} refers to the size of the processor; {\it Quality} evaluates how accurately the quantum processor performs the tasks it was designed or programmed to execute; {\it
Speed} concerns the time required to process a certain task.}
\label{T1}
\vspace{5pt} 
\centering
\begin{tabular}{lccccccc}
\hline
\multicolumn{7}{c}{Quantum Computer Benchmarking Metrics Review} \\
\hline
 Metric & Practical & Repeatable & Reliable & Linear & Consistent & Performance \\
\hline
 Number of Qubits \;             & \checkmark & \checkmark & \checkmark & \checkmark  & \checkmark & Scalability \\
 (Average) Processor Connectivity & \checkmark & \checkmark & $\times$ & $\times$ & \checkmark  & Scalability / Speed \\
Gate Fidelity & \checkmark  &  \checkmark & $\times$ & $-$ & $\times$  & Quality \\
Readout Fidelity & \checkmark  &  \checkmark        & $\times$ & $\times$ & \checkmark  & Quality \\
 (Average) Decoherence Times & \checkmark  &  \checkmark        & $\times$ & $\times$ &  \checkmark  & Quality \\
 (Average) Gate Speed                   & \checkmark & \checkmark & $\times$ & \checkmark & $\times$ & Speed \\
Quantum Volume    & $\times$ & \checkmark & $\times$  & $\times$   & \checkmark  & Scalability / Quality \\
Q-Score & \checkmark & \checkmark & $\times$ & $\times$ & \checkmark & Scalability / Quality \\
 CLOPS  & $\times$ & \checkmark  & \checkmark & $\times$ & $\times$ & Speed \\
Algorithmic Qubits      & $\times$ & \checkmark & $\times$ & $\times$ & \checkmark  & Scalability / Quality\\
Cross Entropy Benchmarking  & $\times$ & \checkmark & \checkmark &  $-$ & \checkmark  & Quality \\
Hellinger distance  & $\times$ & \checkmark & \checkmark  & $-$ &  \checkmark & Quality \\
 Heavy Output Generation       & $\times$ & \checkmark & \checkmark& $-$ &  \checkmark  & Quality \\
$\ell_1$-Norm Distance              & $\times$ & \checkmark & \checkmark &  $-$ & \checkmark  & Quality \\
 Collision volume & $\times$ & \checkmark & \checkmark& $-$ & \checkmark  & Quality \\
\hline
\end{tabular}
\label{tab1}
\end{table}

\section{Roadmap for Quantum Benchmarking}
\label{RoadmapQB}

Building upon earlier sections, our focus shifts towards a more practical exploration of guidelines when designing quantum benchmarks. Our goal in this section is to assist practitioners in evaluating existing benchmarks, making informed choices, and uncovering potential blind spots. To achieve this, we compile practical recommendations and guidelines to steer practitioners in assessing the effectiveness of current benchmarks or when considering the development of new ones. Additionally, we provide a comprehensive view of the benchmarking reporting structure, enabling researchers to convey the true essence of their benchmarks effectively. Finally, we propose the creation of an organization focused on Standard Performance Evaluation for Quantum Computers (SPEQC).

\subsection{Guidelines for the benchmark practitioner}
\label{Guidelines}

Here, we present a set of high-level guidelines to assist benchmark designers in
making informed decisions when creating benchmarks that accurately reflect the
performance of the assessed quantum devices. These guidelines build upon
existing industry standards and take into account insights from classical
benchmarking sources (as discussed in Sec. \ref{Classical Benchmarking}), being
system-independent:

\vspace{.15cm}

\textbf{Guideline 1. Different technological eras necessitate different types of
benchmarks}. In the NISQ era, where there is no dominant architecture for
quantum devices, specification-based benchmarks are particularly relevant. Given
the substantial impact of noise, focusing on synthetic, microbenchmarks or
kernel benchmarks could offer more practical and useful insights. At this early stage of hardware development, it is crucial that benchmarks not only produce comparative scores but also provide detailed information about the performance of individual hardware components. This allows hardware developers to focus on improving their platforms rather than competing solely for benchmark scores. If a benchmark does not provide useful insight to help hardware developers identify which components hinder performance, it does not contribute meaningfully to the improvement of the device. In this context, benchmarks should be understood as tools for guiding technical progress, particularly during the early stages of quantum hardware. This does not mean that other types of benchmarks should not be excluded. In contrast, it is important to design application-oriented benchmarks for algorithms that are relevant to the current era, such as quantum machine learning (QML) algorithms or variational quantum algorithms (VQAs). However, it is less meaningful to prioritize the design of benchmarks that target applications that are not yet feasible with current quantum hardware, such as Shor or Grover algorithms, which are more relevant to future eras of quantum computing. However, recent work has proposed benchmarks based on these algorithms, indicating that this guideline, while seemingly obvious, is not universally followed. As technology
involves, Partially Corrected and Fault-Tolerant quantum computers may
transition into hybrid or kit-based benchmarks. In this Partially Quantum Error Corrected era we will also face the task of developing benchmarks for logical qubits and the quantum error correction methods applied to them, giving rise to a new class of benchmarks focused on fault-tolerant performance and error correction efficacy. Ultimately, for Fault Tolerant
and Fully Functional Quantum Computers eras, application-based benchmarks would
emerge as a reasonable and effective choice. As quantum computers progress,
benchmarks will evolve to reflect this.

\vspace{.15cm}

\textbf{Guideline 2. Benchmarks must uphold the key quality attributes:
relevance, reproducibility, fairness, verifiability, and usability. } In Subsec.
\ref{GoodBenchmark}, we introduced a set of quality attributes initially
suggested as guidelines for classical benchmark designers. We encourage that quantum benchmarks be designed to satisfy these quality attributes. One of our key proposals and recommendations is that any newly defined benchmark should be accompanied by an analysis that clearly identifies which properties it satisfies and which it does not, along with a justification for the relevance or irrelevance of each property in the given context. This practice will help the community understand the appropriate contexts in which these benchmarks or metrics should be applied, as well as highlight key limitations that future developments should aim to address. As an illustrative example, we briefly analyze one of the benchmarks previously discussed in Subsec. \ref{part:qed-c}: the QED-C Benchmark. This benchmark, or benchmark suite, has been continuously updated to incorporate a variety of application-oriented benchmarks for quantum computers. The final output of the benchmark can be interpreted either as a vector comprising the results of multiple specified metrics, or as a collection of individual benchmarks, each with its own set of metrics. Regardless of the interpretation, the resulting analysis remains largely consistent. In this analysis, we treat it as a benchmark suite.
Some components of the suite, such as those based on Shor or Grover algorithms, are not relevant in the current technological era, a point acknowledged in the original guidelines. Others, such as benchmarks based on the Variational Quantum Eigensolver (VQE), remain pertinent. In principle, the benchmarks are reproducible, provided that the corresponding metrics are reported alongside detailed information about all parameters and device calibration. They are also fair, as they do not introduce biases tied to specific hardware architectures or artificial constraints. The methodology is well-defined and established, contributing to the overall verifiability of the suite. However, some benchmarks, such as those involving Hamiltonian simulation, do not scale efficiently to larger qubit counts due to their reliance on classical simulation. This reliance renders them unverifiable for systems with many qubits, as classical resources become insufficient to reproduce or validate the results. Moreover, not all benchmarks in the suite are usable at scale; for example, in cases like Hamiltonian simulation, the required number of measurement shots increases significantly with the number of qubits, making them unsuitable for systems with many qubits.

\vspace{.15cm}

\textbf{Guideline 3. Benchmarks must offer \textit{peak} and \textit{base}
performance results}. The use of \textit{base} and \textit{peak} performance is motivated by the need to enable fair comparison between devices. In addition, fixing the set of allowed optimizations promotes the reproducibility of benchmark results. Base performance is intended to capture a baseline configuration that reflects how a device performs under a common and restricted set of rules, thereby enabling meaningful comparison across architectures. This baseline corresponds to a standard, documented, and restricted configuration that a non-expert user can reasonably reproduce. In contrast, peak performance aims to capture the best achievable performance of a given platform when optimization techniques are employed, allowing innovation without constraining technological progress. Reporting both base and peak results ensures that benchmarks remain comparable while still reflecting the full potential of current quantum processors.

The classical baseline debate with arguments against and
for an extended use of optimization flags can be extended to quantum devices. We
encourage benchmarks practitioners to follow a similar approach as the final
compromise established for SPEC in January 1994: ``Every SPEC CPU benchmark
measurement has to measure performance with a restricted \textit{baseline} set
of flags, and optionally with a more extended \textit{peak} set of
flags"\cite{SPECHPGBenchmarksHistory}. Having both metrics, peak and base,
allows different vendors to share a common baseline for comparison that
represents also the average non expert consumer, while enabling also the
publication of the peak performance with optimised results leveraging expert
knowledge over the architectures. 

In classical computing, compilers are highly advanced and well-established, enabling standardized configuration through flags across different hardware architectures. In contrast, the quantum computing field lacks such standardization: each device typically relies on its own compiler, with no unified approach to transpilation or even to the overall process of circuit preparation. This fragmentation makes it challenging to define a consistent set of configuration flags. Nonetheless, it is possible to identify a set of relevant configuration parameters that could serve as flags in the current NISQ era. It is important to emphasize that each era of quantum computing requires its own definitions of baseline and peak performance. Some example configuration parameters include:

\begin{itemize}

\item \textbf{Classical subroutine simulation:} Allow simulating parts of the circuit or specific subroutines on a classical computer when feasible. For instance, Clifford circuits can be efficiently simulated classically.

\item \textbf{Circuit cancellation:} Permit the elimination of subcircuits that have no effect on the final output, such as a unitary followed immediately by its inverse.

\item \textbf{Circuit equivalence transformations:} Allow transformations of the circuit into another that is equivalent under a well-defined criterion, e.g., preserving the final quantum state. For example, conjugating a Pauli gate with a Clifford circuit results in another Pauli gate.

\item \textbf{Approximate circuit transformation:} Permit transformations that yield a circuit producing a result $\epsilon$-close (under a defined metric) to the original target outcome. An example includes rearranging gates that do not strictly commute to obtain a similar final state.

\item \textbf{Error mitigation:} Specify whether error mitigation techniques are allowed, and if so, define which classes of techniques (e.g., zero-noise extrapolation, probabilistic error cancellation) are permissible.

\item \textbf{Measurement strategies:} Include flags for measurement techniques, which are especially relevant in the NISQ era due to the significant error contribution from the measurement process. Examples include changing the measurement basis to reduce the number of shots (e.g., grouping commuting Pauli observables) or using advanced techniques like classical shadows to efficiently estimate multiple observables.

\item \textbf{Hardware optimization:} Techniques that depend on the specific characteristics of the quantum device, such as qubit fidelity, coherence times, gate connectivity, native operations, among others. For example, insert the minimal number of SWAP gates when qubits are not directly connected, selecting the lowest-cost path according to the device connectivity.

\end{itemize}

\vspace{.15cm}

\textbf{Guideline 4. Benchmarks metrics must uphold the key quality attributes:
practical, repeatable, reliable, and consistent. They should also preferably be
independent and linear.} As discussed in Sec. \ref{Classical Benchmarking}, for
a metric to be useful in build a good benchmark it must present attributes that
correlate later with quality benchmark attributes. Additionally, while linearity
and independence are desirable metric attributes, they are not strictly required
for effectively capturing the performance of a quantum device. Regarding benchmarks, we encourage that each proposal of a new metric includes an analysis of the properties it satisfies as well as those it does not.
\vspace{.15cm}

\textbf{Guideline 5. Do not rely solely on a single benchmark to measure
performance.} While benchmarks serve as valuable tools for modeling specific
aspects of the behavior of a quantum computer, relying solely on individual
benchmarks may result in a lack of depth. The risk of spreading benchmarks too
thin clearly arises when attempting to cover the entirety of the behavior of a
device space with a single metric. For example, metrics such as Quantum Volume or Cross-Entropy Benchmarking (XEB) can provide a general overview of a quantum computer's performance. However, to obtain a more comprehensive understanding, it is essential to incorporate additional, more specific metrics that assess individual characteristics of the system.
Different behaviors often require distinct
testing and experimentation setups, and consequently, an approach based on
benchmark suites is recommended. Benchmark suites provide a more comprehensive
assessment, enabling practitioners to really grasp how the computers perform in
different situations.

\begin{figure}
    \centering
    \includegraphics[width=\linewidth]{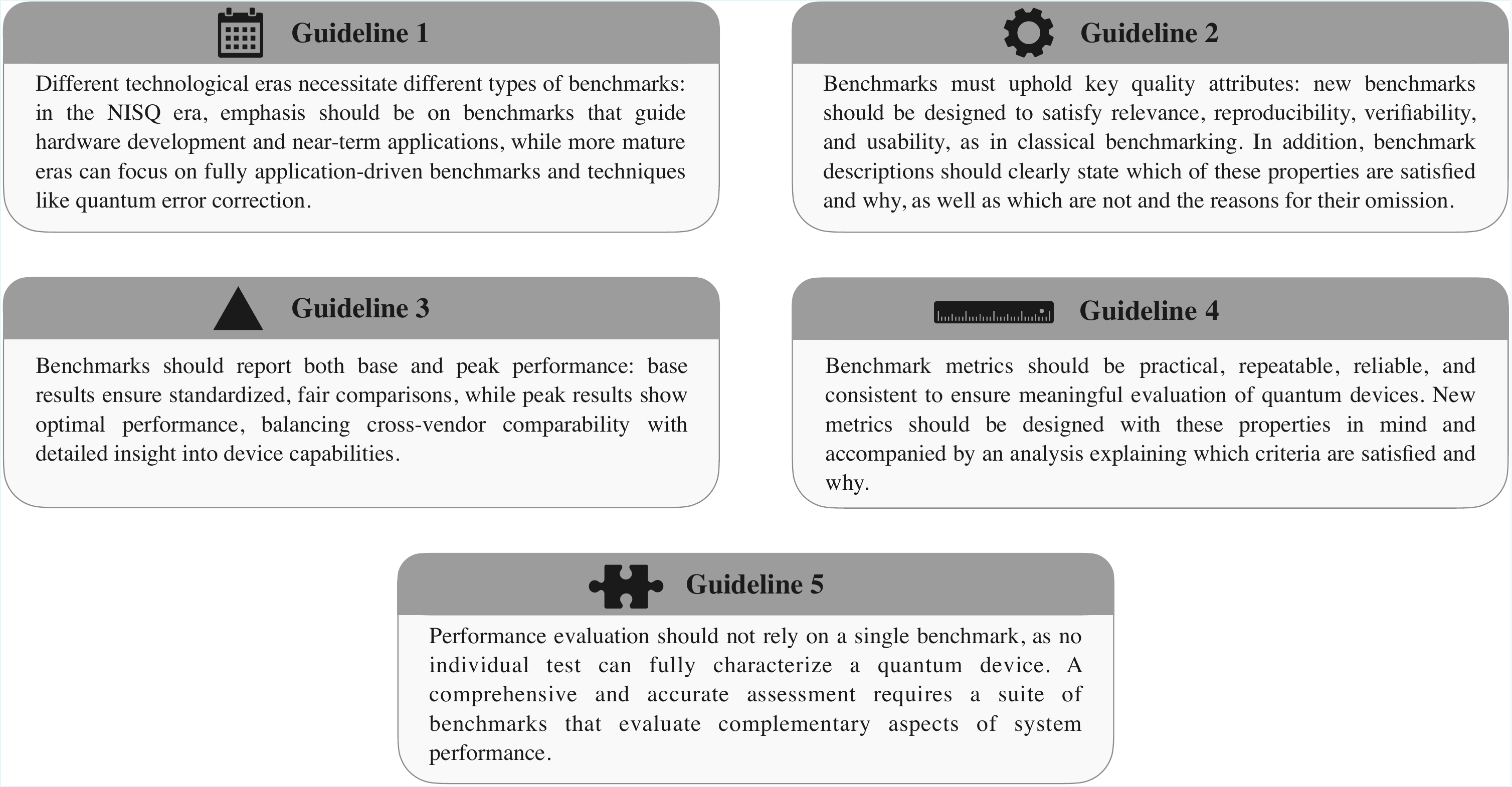}
    \caption{Summary of the proposed roadmap for quantum benchmarking. The figure highlights five guidelines addressing (1) the need for era-dependent benchmarking goals, (2) key benchmark quality attributes, (3) the reporting of base and peak performance, (4) desirable properties of benchmark metrics, and (5) the importance of using complementary benchmark suites rather than relying on a single test.}
    \label{fig:guidelines}
\end{figure}
\vspace{.15cm}

\subsection{Benchmarking structure}
\begin{figure}[t!]
    \centering 
    \includegraphics[width=0.50\columnwidth]{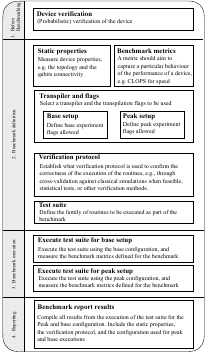}
    \caption{{\bf Quantum benchmarking structure}. We introduce a step-wise structure to benchmarking, from pre-benchmark steps for verification of the device, defining the benchmarking and executing it, up to the reporting stage. }
    \label{Fig:BenchmarkingProcess}
\end{figure}

Beyond adhering to the guidelines outlined in the previous section, we present a
high-level step-wise benchmarking structure that encompasses the primary aspects
benchmarks practitioners must consider. We anticipate that this general schema
will promote a standardized approach to benchmarking, facilitating streamlined
comparison efforts among various alternatives in the future, while also ensuring
consistency in meeting a shared quality standard. This step-wise schema is
detailed in Fig. \ref{Fig:BenchmarkingProcess}.

Device verification is a crucial process to ensure that the device is working
properly. Just as vendors in classical computing certify the validity of their
products before consumer release, similar steps must be taken in the quantum
computing field. However, the techniques for verification differ due to the
fundamentally different nature of quantum computers, rooted in quantum
mechanics. Since we are in NISQ era, the concepts of
verification and benchmarking overlap because we can not ensure that a quantum
computer will work properly. An example of device verification method is the
cross-entropy benchmarking (XEB), which has been employed to verify quantum
processors based on superconducting circuits \cite{Neill_2018,Arute2019}.

After the device verification, we proceed with the benchmark definition. First,
we define the benchmark or benchmark suite that we will execute to measure the
performance of the quantum processor for a certain task and define static or
non-static metrics that capture the behavior of interest of the device. To
execute the benchmark, users select a compiler to transform it into a quantum
circuit, optimizing and converting it into an executable version on the quantum
processor, considering constraints such as native supported quantum operations
and qubits connectivity. Additionally, compilation rules to be allowed must be
defined for both base setup and peak setup to ensure we have a baseline to
compare benchmark results, as well as allowing the comparisons for peak
performance when custom or expert modifications are applied. Finally, it must be
established which verification protocol will be used to confirm that the
benchmark has been correctly executed.

Once the benchmark has been defined, we continue with its execution. We execute
the benchmark suite defined using the base and setup configurations, and measure
the metrics defined for the benchmark. It is advisable to run the benchmarks
multiple times and take the average of the results to reduce the impact of
errors. Sometimes, the benchmark suite may not execute as expected due to a
variety of reasons. Thus, it is important to verify the benchmark results using
the verification protocol to ensure the benchmark is executing properly.

After completion of the benchmark execution and verification, we proceed to
report the results. We must elaborate reports compiling all results from the
execution of the test suites for the peak and base configuration.  Additionally,
it should include the static and non-static metrics, the verification protocol,
the configuration used for peak and base executions and execution process, so
that others users can reproduce and verify your results. This report should be
clear and contain as much information as possible so that users know how the
computers are performing and can identify areas where the computers are
underperforming.

Note that while the proposed step-wise approach for quantum benchmarking bears
resemblances to the classical benchmarking paradigm, a closer examination
reveals significant distinctions. In particular, the specific tasks required to complete each step are notably distinct. 

For example, in quantum computing, we are still figuring out how to define the
transpilation process and what configuration/flags are available, therefore it
is not yet as developed as the compilation techniques used in classical
computing. Also, unlike transpilers that maintain the abstraction level,
compilers inherently involve abstraction level transformations from high-level
programming languages to assembly primitives. Nonetheless, both face the
challenge of arranging a primitive set of instructions optimally for the device
they target. Still, because of the different goals and rules involved in
compilation and layout rearrangement, unfortunately, there is no direct
transferability of the knowledge.

The structure outlined provides a roadmap for pinpointing areas that require
further investigation to clarify the future of quantum computer benchmarking.
Researchers and benchmark practitioners are actively exploring each of the
tasks, aiming to enhance our understanding and methodologies in this field.
Below, we outline the primary challenges and unresolved issues that must be
addressed to establish a more robust and comprehensive benchmarking process for
quantum computers:

\begin{itemize}
    \item \textbf{Device and routine execution verification}. Post-NISQ era, a
    significant challenge we must address, which is becoming increasingly
    apparent with advancements in available quantum computers even within NISQ,
    is verifying the correctness of the results they produce. As mentioned
    earlier in Sect. \ref{Quantum Benchmarking}, we know that the problem to
    verify if every language in BQP admits an interactive proof where the prover
    is in BQP too, and the verifier is in BPP is still opened
    \cite{AaronsonPrize}. This question has been thoroughly examined within the
    field, even extending to inquiries about the falsifiability of quantum
    mechanics as a theory \cite{aharonov2012quantum}. Still, it is not the only
    approach explored for verifiability so far in the field, with other
    protocols explores with single-provers and multi-provers either quantum or
    classical \cite{Gheorghiu2018}.
    
    In practice, verification in quantum computing depends both on the computational era—such as the NISQ or post-NISQ era—and on the specific aspect of the quantum computer that is being verified. In the NISQ era, verification remains feasible for small system sizes through cross-validation with classical simulations, which compare the results generated by quantum devices with exact solutions for a small number of qubits or with approximate classical methods applicable only in restricted regimes. In addition, statistical tests can be employed to assess whether experimentally estimated quantities exceed predefined thresholds with a given level of confidence, thereby providing probabilistic guarantees even when exact verification is not possible. As quantum processors scale beyond the NISQ regime, verification becomes substantially more challenging. In the post-NISQ era, building confidence in large-scale quantum computations relies on complementary forms of indirect evidence. These include ensemble-based analyses across related problem instances, internal consistency checks based on properties expected to be preserved by the computation, robustness tests under controlled perturbations, validation against analytical results where available, and cross-verification across different quantum devices \cite{hashim2024,Gheorghiu2018}.

    Finally, It is also important to distinguish between different notions of verification. On the one hand, verification may refer to assessing whether a given quantum circuit has been executed correctly on a device, which is typically addressed through measures related to process fidelity, implementation accuracy, or benchmarking techniques. On the other hand, verification may refer to establishing that a computation achieves a genuine quantum advantage over classical methods, which constitutes a substantially harder and more fundamental problem rooted in computational complexity. While the former is an essential component of benchmarking and device characterization, the latter generally lies beyond the scope of experimental verification. Acknowledging this distinction is crucial for the correct interpretation of benchmarking results.

    \item \textbf{Definition of \textit{good} metrics}.  
    In this study, we have introduced in Subsec. \ref{GoodMetric} the quality
    attributes that \textit{good} metrics should ideally meet. Additionally, we
    have examined various proposals put forth by researchers in the field in
    Subsec. \ref{Reviewing}. However, we contend that further work is necessary
    to precisely define the \textit{relevance} of the metrics. In other words,
    we need to determine which aspects of the devices behavior space are being
    captured by these measurements. Establishing a correlation between metrics
    and their actual relevance to performance is crucial for obtaining diverse
    perspectives of the device, and a thorough assessment. Throughout the NISQ era, prioritizing metrics focused on scalability, accuracy, and quality can better guide the development of robust and practically useful quantum devices. While gate speed remains a relevant and informative metric—especially given the variability in gate durations across different technological platforms—we consider it complementary to those assessing overall system performance and reliability. In our view, scalability, accuracy, and quality represent more immediate challenges for current quantum computing architectures. Therefore, speed-related metrics are valuable, but should be regarded as secondary to those that capture the system’s computational potential and robustness. However, speed may become a
    necessary performance indicator as we transition towards the Fault Tolerant
    era or implement additional partial quantum error correction methods. 
    
    \item \textbf{Standarise transpilation and compilation process}. While we
    have discussed in detail the problematic regarding the transpilation, which
    can be mapped to the idea of how to optimise the intermediate representation
    (IR) language descriptor for quantum programs (e.g. OpenQASM
    \cite{cross2017open} or QASM \cite{1580386}), we have not detailed the
    mapping into the physical layer. This aspect of the compilation and
    transpilation funnel relates to the \textit{Quantum Physical Operations
    Language} (QPOL) \cite{1580386}. Due to the lack of a standard architecture
    for the development of the quantum technologies and the complex nature of
    simplifying the intermediate representation (IR) of circuits, much of the
    research focus has revolved around this aspect. In the future, we might have
    to include the exploration of QPOL optimization techniques through standard
    compilers within the industry. Even today, transpilation alone is witnessing
    a surge of initiatives aiming to delve deeper into its complexities.
    Techniques for gate decomposition, routing, and mapping are continuously
    evolving, alongside other approaches have further explored ideas such as
    parameterized circuit instantiation where a feedback loop is allowed
    \cite{9951320}. No standards or rules are defined to what is the
    \textit{fair} use of these techniques for quantum devices. From the
    classical computing benchmarking past, we know good examples of techniques
    that are not allowed to encourage a fair competition of benchmark results.
    For instance, the definition of assertion flags is not allowed, i.e. flags
    that assert a certain property of the program, one that may not hold for
    other routines. Another example of a restriction to the configuration of a
    classical benchmark execution is that no more than two compilation passes
    without processing in between are permitted \cite{SPECHPGBenchmarksHistory}.
    We believe on the quantum computing field there is still a need to further
    explore the fair usage of optimization routines and flags on transpilations
    to standarise the comparison of benchmark results for different devices,
    resembling these previous efforts on classical benchmarking. This debate
    extends too to the process of establishing peak and base configurations.
    
    \item \textbf{Matching routines to relevant regions of the behaviour space}.    
    Even though we are still in the early ages of quantum computing, there is a
    huge investment to find and identify real-world applications that can
    benefit from this technology. This search is particularly relevant for
    defining which regions of the behavior space of a quantum device need to be
    evaluated. This effort needs to be recurrent, to prevent us from falling
    prey to the Goodhart's law as warned before. We firmly believe an active
    effort to keep uncovering what aspects of the computational device are
    useful is required to keep modelling new benchmarks accordingly. These new
    discoveries will make some of our current benchmarks obsolete, while opening
    the door for future benchmarks to be created. As Hennessy and Patterson
    pointed out in their classic book on computing architecture, it is
    fallacious to assume that benchmarks retain their validity indefinitely
    \cite{Hennessy2011-kt}. 
    
\end{itemize}

\begin{sectionsummary}{Key points of the section: Roadmap for Quantum Benchmarking}
\begin{itemize}\setlength\itemsep{2pt}

  \item \textbf{What benchmarking is:} In quantum computing, a benchmark is a test or test suite used to compare quantum processors by measuring performance under a specified workload.
  \item \textbf{What performance means:} Performance is inherently multidimensional and context-dependent, involving aspects such as execution time, throughput, resource usage, accuracy, energy consumption, and scalability.
  \item \textbf{Central challenge:} The choice of workload is critical—benchmarks only measure performance relative to the behaviors they exercise, and unrepresentative workloads lead to misleading results.
  \item \textbf{Good benchmark principles:} Useful benchmarks aim to be relevant to real use cases, reproducible, fair across systems, verifiable, and usable in practice, though these properties often involve trade-offs.
  \item \textbf{Benchmark diversity:} Benchmarks can differ by workload strategy (fixed-work, fixed-time, or variable), by implementation level (specification- vs.\ kit-based), and by realism (synthetic tests to full applications).
  \item \textbf{Role of metrics:} Benchmarks rely on performance metrics, which should be practical, repeatable, reliable, and consistent; no single metric captures all aspects of performance.
  \item \textbf{Interpretation risks:} Normalization, aggregation, and ratio-based comparisons can be manipulated, reinforcing that performance cannot be reduced to a single number.
  \item \textbf{Main conclusion:} There is no universal benchmark; meaningful evaluation requires multiple complementary benchmarks and careful interpretation aligned with the evaluation objective.
  
\end{itemize}
\end{sectionsummary}

\subsection{Standard Performance Evaluation for Quantum Computers (SPEQC)}
\label{SPEQC}

As we have discussed in this article, assessing the performance of a quantum
processor presents a challenge without a straightforward solution. This
underscores the risks of relying solely on a single benchmark or, even worse,
relying on a single metric. Therefore, it is essential for us to collectively
establish a standardized set of \textit{good} benchmarks. These benchmarks would
facilitate effective evaluation and comparison of different quantum processors
for both users and manufacturers.

While the development of quantum computing standardized benchmarks is a valuable and necessary effort, it is important to recognize that a standardization may introduce certain risks. One of these risks is that early standardization of benchmarking could unintentionally hinder the advancement of promising technologies that are still in earlier stages of development. This risk arises when benchmarking criteria reflect only the capabilities of more mature quantum hardware, thereby creating implicit performance expectations that emerging technologies cannot meet. As a result, research and investment may become concentrated on a narrow subset of quantum platforms, reducing diversity and potentially overlooking technologies with long-term advantages. To avoid this outcome, it is essential that benchmarking frameworks be flexible, inclusive, and responsive to the evolving technological landscape of quantum computing. For example, benchmarks could be designed in a modular way to provide flexibility, adopting layered structures, such as synthetic, microbenchmarks, kernel, hybrid, kit-based and application-level benchmarks, that can be tailored to the current stage of the technology under evaluation (NISQ, PQEC, FTQC, FFQC). To move in this direction, it would first be necessary to classify and characterize both existing and future benchmarks, and subsequently to accurately determine the technological stage of the quantum platform under evaluation. This would enable the application of appropriate criteria and fair evaluation based on the hardware level of maturity.

Building on these ideas and drawing inspiration from the establishment of the Standard Performance Evaluation Corporation (SPEC) in classical processor benchmarking,
we propose the creation of an organization dedicated to Standard Performance
Evaluation for Quantum Computers (SPEQC) as a flexible benchmarking framework that adapts to the different stages and technologies of quantum hardware development. Rather than enforcing a fixed set of metrics or algorithms, SPEQC should be designed to accommodate a diverse range of benchmark types. It is worth noting that SPEQC should not be limited to application or full-stack benchmarks. Instead, it should also incorporate low-level methodologies, including synthetic, microbenchmark, and kernel-based approaches, thereby providing a comprehensive framework for evaluating both individual hardware components and overall system performance across diverse quantum architectures.

SPEQC should be a non-profit organization dedicated to establishing criteria for
evaluating the performance of quantum computers, as well as developing and
distributing a standardized set of tests to be used as benchmarks. Furthermore,
SPEQC should provide the industry with clear specifications to assist benchmark
designers in creating assessments that accurately reflect the performance of
quantum devices. In general, these specifications should follow the guidelines
for the benchmarks introduced in Subsec. \ref{Guidelines}.

Performance evaluation standards provide a common framework for assessing the
performance of quantum computers, reducing the potential for bias in the
evaluation process. When evaluation criteria are clearly defined and applied
uniformly, it becomes easier to compare different results and objectively
determine the performance level of the devices. Additionally, performance
evaluation standards help maintain consistency in the evaluation process,
ensuring that all quantum devices are evaluated using the same criteria. This
consistency is essential for guaranteeing fairness and avoiding any perception
of favoritism towards a particular architecture.

\begin{figure}[ht]
    \centering 
    \includegraphics[width=0.75\columnwidth]{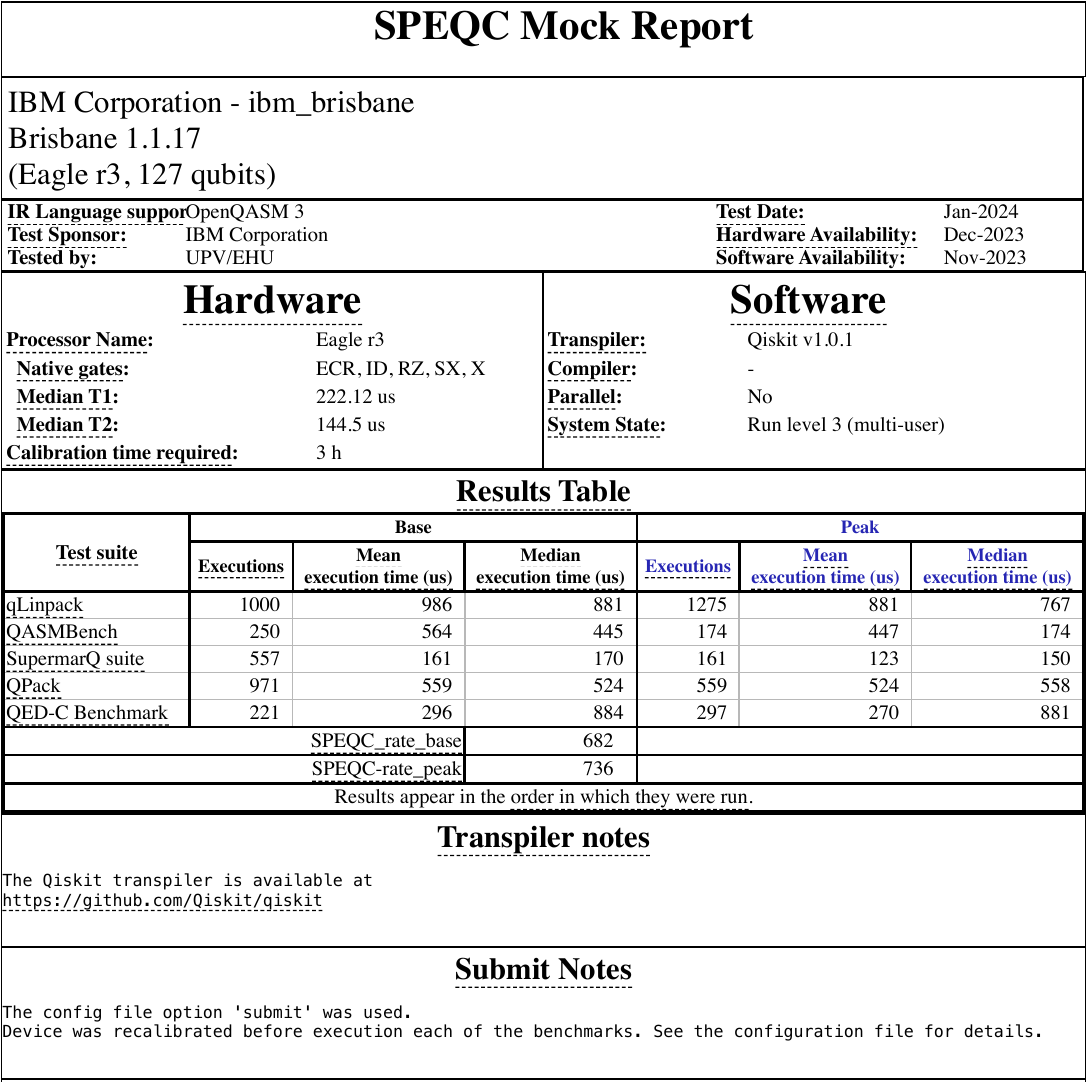}
    \caption{{\bf Example of SPEQC report}. Report inspired by the \href{https://www.spec.org/cpu2017/}{SPEC CPU® 2017}, none of the metrics are reflecting actual measurements of the devices or real properties, their usage is limited to exemplify how the report would look. More details can be added to the report file in the form of sections below Submit Notes.}
    \label{Fig:Speqc_mock}
\end{figure}

Establishing standards for execution, compilation, and transpilation, as well as
rules for reporting results and standards, would ensure consistency, clarity,
credibility, and, most importantly, the ability to reproduce performance
results. Fig. \ref{Fig:Speqc_mock} shows an example of how quantum processor
performance results could be reported. The report would offer detailed
information about the hardware and software, including the processor name,
native gates, transpiler, and other relevant details. Furthermore, the results
tables provide the performance of the quantum processor when executing different
benchmarks. We emphasize that this results report is an example and may contain
more information. These reports should be accessible to provide consumers with
transparent information about manufacturers product performance, facilitating
informed decision-making across the industry.

The development of benchmarks is a complex task and requires making significant
decisions that must satisfy several design objectives simultaneously. In other
words, when designing benchmarks, it is necessary to balance different
objectives that are important for the design and functionality of the benchmark.
These objectives may include ease to use, relevance, and fairness,
among others. One of the most relevant aspects to highlight is the need to
ensure that the benchmarks developed are fair. To achieve this, SPEQC should
actively involve and collaborate with a diverse range of stakeholders. This
would involve participation representatives from  various sectors, including the
scientific community, industry, public administration, and hardware and software
vendors. A benchmark developed with the participation of all stakeholders is
considered fairer than one designed exclusively by a single company. Although
involving all stakeholders may not be the most efficient method for developing a
benchmark, consensus among multiple stakeholders will make the benchmark fair.
Consequently, SPEQC should adopt an inclusive membership policy, whereby any
company or organization with an interest in advancing the field of quantum
computing is encouraged and welcomed to join the initiative. This inclusive
approach would ensure that the development and implementation of benchmarks are
reflective of the broader interests and needs of the quantum computing
community, thereby maximizing their impact and relevance in the industry. By
promoting an open and collaborative environment for debate, the exchange of
knowledge, perspectives, and best practices is facilitated, driving innovation
and progress in the field of quantum computing.

\section{Conclusion}
\label{Conclusions}

In summary, we have proposed guidelines to assist benchmark developers in
creating fair and good benchmarks. Drawing from extensive experience in the
field of classic benchmarks, we have defined the quality attributes that must be
met for a quantum benchmark to be considered good. It is important to note that
meeting all these attributes simultaneously can be challenging. We recommend
using benchmark suites that include different tests to assess the performance of
quantum computers in various tasks. These suites offer a more comprehensive
evaluation, allowing professionals and users to truly understand how quantum
computers operate in different work scenarios.

Additional, drawing from past lessons, we critically reviewed the existing
methods and strategies proposed for quantum benchmarking. We identified
inconsistencies in the terminology and concepts used in this field. To address
this, we propose adopting a standardized vocabulary for quantum computer
benchmarking. This common language will help structure our collaborative efforts
more clearly, enabling more accurate measurement of quantum performance.

Along with proposing guidelines for benchmark designers, we have introduced a
benchmarking structure that includes the essential elements that benchmark
professionals should consider. This structure provides a clear guide on the
necessary steps to ensure a reliable and fair comparative evaluation process.
Identifying the essential elements helps avoid biases and establishes quality
standards for evaluating quantum devices. Furthermore, this structure highlights
areas that require more attention to advance the field of quantum computing
benchmarking. 

Standardization in the evaluation of quantum device performance is essential for
fair and objective assessments, ensuring the quality and reliability of test
results. To this end, we propose the creation of SPEQC, a nonprofit organization
dedicated to establishing clear criteria for evaluating quantum computers. By
defining these criteria, biases and subjectivities are eliminated, resulting in
more accurate and reliable evaluations. Our focus on advancing the
standardization of quantum device performance evaluation aims to drive progress
in quantum computing technology in the right direction, while acknowledging the
risks of falling prey to Goodhart's Law. For this reason, we expect that our
benchmarking framework, along with continuous efforts from the community, will
help us stay on the right track, always vigilant for potential pitfalls in our
approaches.

\section{Acknowledgments}
We thank T. Monz, A. Frisk-Kockum, Z. Zimboras, and F. Wilhelm-Mauch for their
helpful discussions and valuable comments, and for carefully reading our
article. The authors acknowledge financial support from OpenSuperQ+100 (Grant
No. 101113946) of the EU Flagship on Quantum Technologies, as well as from the
EU FET-Open project EPIQUS (Grant No. 899368), also from Project Grant No.
PID2021-125823NA-I00 and Spanish Ram\'on y Cajal Grant No. RYC-2020-030503-I
funded by MCIN/AEI/10.13039/501100011033 and by “ERDF A way of making Europe”
and “ERDF Invest in your Future,” this project has also received support from
the Spanish Ministry of Economic Affairs and Digital Transformation through the
QUANTUM ENIA project call –Quantum Spain, and by the EU through the Recovery,
Transformation and Resilience Plan– NextGenerationEU within the framework of the
Digital Spain 2026 Agenda. We acknowledge funding from Basque Government through
Grant No. IT1470-22, and the ELKARTEK project KUBIBIT - kuantikaren berrikuntzarako ibilbide teknologikoak (ELKARTEK25/79), and the IKUR Strategy under the collaboration agreement between Ikerbasque Foundation and BCAM on behalf of the Department of Education of the Basque Government. This work has been supported by Basque Government through the BERC 2022-2025 program.


\begin{thebibliography}{2}
\bibitem{6448963} Dongarra J, Martin J L and Worlton J 1987 Computer benchmarking: paths and pitfalls \textit{IEEE Spectrum} \href{https://doi.org/10.1109/MSPEC.1987.6448963}{\textbf{24} 38--43}

\bibitem{HowNotToLieWithStatistics} Fleming P J and Wallace J J 1986 How not to lie with statistics: the correct way to summarize benchmark results \textit{Communications of the ACM} \href{https://doi.org/10.1145/5666.5673}{\textbf{29} 218--221}

\bibitem{Jacob1995NotesOC} Jacob B and Mudge T N 1995 Notes on calculating computer performance Advanced Computer Architecture Lab (tech. report) \url{https://tnm.engin.umich.edu/wp-content/uploads/sites/353/2021/06/1995_Notes_on_calculating_computer_performance.pdf}

\bibitem{CharacterizingComputerPerformanceWithASingleDigit} Smith J 1988 Characterizing computer performance with a single number Communications of the ACM \href{https://doi.org/10.1145/63039.63043}{\textbf{31} 1035--1039}

\bibitem{HowToBuildABenchmark} Kistowski J \textit{et al} 2015 How to build a benchmark Proceedings of the 6th ACM/SPEC International Conference on Performance Engineering \href{https://doi.org/10.1145/2668930.2688819}{333--336}

\bibitem{Eigenmann2001nl} Eigenmann R 2001  Performance evaluation and benchmarking with realistic applications The MIT Press

\bibitem{Lilja2005o} Lilja D J 2005 Measuring computer performance \textit{Cambridge University Press}

\bibitem{Jain1991-xw} Jain R 1991 The art of computer systems performance analysis \textit{John Wiley \& Sons}

\bibitem{Kounev2020mh} Kounev S, Lange K-D and von Kistowski J 2020 Systems benchmarking  Springer

\bibitem{leboudec2010performance} Le Boudec J-Y 2010 Performance evaluation of computer and communication systems  EPFL Press

\bibitem{Feitelson2015} Feitelson D G 2015 Workload modeling for computer systems performance evaluation Cambridge University Press

\bibitem{Sim_BenchmarkingToAdvanceResearch} Sim S E, Easterbrook S and Holt R C 2003 Using benchmarking to advance research: A challenge to software engineering Proc. 25th International Conference on Software Engineering \href{https://ieeexplore.ieee.org/document/1201189}{74--83}

\bibitem{Shor1994} Shor P W 1994 Algorithms for quantum computation: discrete logarithms and factoring Proc. 35th Annual Symposium on Foundations of Computer Science (FOCS) \href{https://doi.org/10.1109/SFCS.1994.365700}{124--134}

\bibitem{WhetstoneBook} Wichmann B 1973 Algol 60 compilation and assessment \textit{Academic Press, Studies in Data Processing} \href{https://doi.org/10.1002/spe.4380040412}{\textbf{4} 241--255}

\bibitem{DhrystonePaper} Weicker R P 1984 Dhrystone: a synthetic systems programming benchmark \textit{Communications of the ACM} \href{https://doi.org/10.1145/358274.358283}{\textbf{27} 1013--1030}

\bibitem{Hennessy2012} Hennessy J L and Patterson D A 2012 Computer architecture: a quantitative approach 5th ed. Morgan Kaufmann

\bibitem{goodhart1975problems} Goodhart C 1984 Problems of Monetary Management: The UK Experience  In: Monetary Theory and Practice. Palgrave

\bibitem{SkadronChallengesInComputerArchitectureEvaluation} Skadron K \textit{et al} 2003 Challenges in computer architecture evaluation \textit{Computer} \href{https://ieeexplore.ieee.org/document/1220579}{\textbf{36} 30--36}

\bibitem{osti_1775058} Blume-Kohout R, Nielsen E, Proctor T \textit{et al} 2020 Not all benchmarks are created equal Workshop: Quantum Protocols—Testing \& Quantum PCPs 

\bibitem{osti_1648785} Blume-Kohout R 2019 Metrics \& benchmarks for digital quantum computing May Sandia National Laboratories (SNL-NM)

\bibitem{osti_1639885} Blume-Kohout R 2019 Performance standards for quantum computing May Sandia National Laboratories (SNL-NM)

\bibitem{osti_1594677} Blume-Kohout R and Young K 2018 Quantum computing metrics and benchmarking frameworks: not yet Rebooting Computing Workshop on Quantum Metrics \& Benchmarks November Sandia National Laboratories (SNL-NM)

\bibitem{quantumComputingSummitWhitepaperAtlanta} Gomes L 2018 Quantum Computing Summit (white paper) \url{https://quantum.ieee.org/images/files/pdf/ieee-quantum-summit-white-paper-2018.pdf}

\bibitem{StandrdForQuantumComputingPerformanceIEEE} IEEE Computer Society 2023 Standard for quantum computing performance metrics \& performance benchmarking \textit{IEEE Standards Association}

\bibitem{amico2023defining} Amico M, Zhang H, Jurcevic P \textit{et al} 2023 Defining standard strategies for quantum benchmarks (arXiv:\href{https://doi.org/10.48550/arXiv.2303.02108}{2303.02108})

\bibitem{BlumeKohout2019} Blume-Kohout R and Young K 2019 Metrics and benchmarks for quantum processors: state of play \textit{Sandia Report} \href{https://doi.org/10.2172/1493362}{\textbf{1493362}}

\bibitem{specglossary} Standard Performance Evaluation Corporation Glossary \url{https://www.spec.org/spec/glossary/}

\bibitem{PP} (Context) We focus in this article on the study of benchmarking related to physical devices or processors, even though interesting debates are held by the community in other benchmarking fields such as algorithms benchmarking \cite{BenchmarkingOptimization,TestingHeuristics}.

\bibitem{Zeigler2018fg} Zeigler B P, Muzy A and Kofman E 2018 Theory of modeling and simulation 3rd ed. Academic Press

\bibitem{SimulationModelingAndAnalysisLawAverill} Law A 2023 Simulation modeling and analysis 6th ed. McGraw-Hill Education

\bibitem{Wu2021} Wu X \textit{et al} 2021 SeQUeNCe: a customizable discrete-event simulator of quantum networks \textit{Quantum Sci. Technol.} \href{http://dx.doi.org/10.1088/2058-9565/ac22f6}{\textbf{6} 045027}

\bibitem{Coopmans2021} Coopmans T, Knegjens R, Dahlberg A \textit{et al} 2021 NetSquid: a network simulator for quantum information using discrete events \textit{Communications Physics} \href{https://doi.org/10.1038/s42005-021-00647-8}{\textbf{4} 157}

\bibitem{Tripathi} Tripathi V, Kowsari D, Saurav K \textit{et al} 2025 Benchmarking quantum gates and circuits \textit{Chemical Reviews} \href{http://dx.doi.org/10.1021/acs.chemrev.4c00870}{\textbf{125} 5745--5775}

\bibitem{DeRaedt2005} De Raedt K, De Raedt H and Michielsen K 2005 Deterministic event-based simulation of quantum phenomena \textit{Computer Physics Communications} \href{http://dx.doi.org/10.1016/j.cpc.2005.04.012}{\textbf{171} 19--39}

\bibitem{Willsch2020} Willsch M \textit{et al} 2020 Discrete-event simulation of quantum walks \textit{Frontiers in Physics} \href{http://dx.doi.org/10.3389/fphy.2020.00145}{\textbf{8} 145}

\bibitem{ComputerResearchWorkload} Eeckhout L, Vandierendonck H and De Bosschere K 2003 Designing computer architecture research workloads \textit{Computer} \href{https://doi.org/10.1109/MC.2003.1178050}{\textbf{36} 65--71}

\bibitem{ArtificialWorkloadDesign} Ferrari D 1984 On the foundations of artificial workload design \textit{Proc. ACM SIGMETRICS} \href{https://doi.org/10.1145/1031382.809309}{\textbf{12} 8--14}

\bibitem{GustafsonAndSnell_HINT} Gustafson J and Snell Q 1995 HINT: a new way to measure computer performance \textit{Proc. 28th Hawaii International Conference on System Sciences} \href{http://dx.doi.org/10.1109/HICSS.1995.375519}{\textbf{1995} 392--401}

\bibitem{Henning_SpecCpu2000} Henning J L 2000 SPEC CPU2000: measuring CPU performance in the new millennium \textit{Computer} \href{https://doi.org/10.1109/2.869367}{\textbf{33} 28--35}

\bibitem{Huppler_TheArtOfBuildingGoodBenchmark} Huppler K 2009 The art of building a good benchmark \textit{Lecture Notes in Computer Science} \href{https://doi.org/10.1007/978-3-642-10424-4_3}{\textbf{5895} 18--38}

\bibitem{cpu2017} Standard Performance Evaluation Corporation 2017 SPEC CPU2017 Docs \url{https://www.spec.org/cpu2017/Docs/}

\bibitem{P} (Definition) In a ratio game, we are comparing devices performance using ratios, where the denominator is called the \textit{base}.

\bibitem{Mashey2004} Mashey J 2004 War of the benchmark means \textit{ACM SIGARCH Computer Architecture News} \href{https://doi.org/10.1145/1040136.1040137}{\textbf{32} 1--14}

\bibitem{SLALOMBenchmark} Gustafson J \textit{et al} 1991 The design of a scalable, fixed-time computer benchmark \textit{Journal of Parallel and Distributed Computing} \href{https://doi.org/10.1016/0743-7315(91)90008-W}{\textbf{2} 388--401}

\bibitem{Gustafson1992} Gustafson J L 1992 The consequences of fixed time performance measurement \textit{Proc. 25th Hawaii International Conference on System Sciences} \href{http://dx.doi.org/10.1109/HICSS.1992.183285}{\textbf{1992} 113--124}

\bibitem{Isoefficiency} Grama A Y, Gupta A and Kumar V 1993 Isoefficiency: measuring the scalability of parallel algorithms and architectures \textit{IEEE Parallel \& Distributed Technology} \href{https://doi.org/10.1109/88.242438}{\textbf{1}(3) 12--21}

\bibitem{ReflectionsOnCreationParallelBenchmarkSuite} Van Voorst B \textit{et al} 2006 Reflections on the creation of a real-time parallel benchmark suite \textit{Parallel and Distributed Processing, Lecture Notes in Computer Science} \href{https://doi.org/10.1007/BFb0098014}{\textbf{1586} 390--397}

\bibitem{OverviewOfCommonBenchmarks} Weicker R P 1990 An overview of common benchmarks \textit{Computer} \href{https://doi.org/10.1109/2.62094}{\textbf{23}(12) 65--75}

\bibitem{IsaWars} Blem E \textit{et al} 2015 ISA wars: understanding the relevance of ISA being RISC or CISC to performance, power, and energy on modern architectures \textit{ACM Transactions on Computer Systems} \href{https://doi.org/10.1145/2699682}{\textbf{33} 1--34}

\bibitem{dongarra1979linpack} Dongarra J J, Bunch J R, Moler C B and Stewart G W 1979 LINPACK users' guide SIAM

\bibitem{Dongarra2003TheLB} Dongarra J J, Luszczek P and Petitet A 2003 The LINPACK benchmark: past, present and future Concurrency and Computation: Practice and Experience \href{https://doi.org/10.1002/cpe.728}{\textbf{15} 803--820}

\bibitem{top500} TOP500 List \url{https://www.top500.org/}

\bibitem{CriticsTop500} Kramer W T C 2012 Top500 versus sustained performance: The top problems with the Top500 list—and what to do about them \textit{Proc. PACT '12} \href{https://doi.org/10.1145/2370816.2370850}{\textbf{2012} 223--230}

\bibitem{PerformanceModelingHPCG} Marjanović V, Gracia J and Glass C 2015 Performance modeling of the HPCG benchmark \textit{Performance Modeling, Benchmarking, and Simulation (PMBS 2014), Lecture Notes in Computer Science} \href{https://doi.org/10.1007/978-3-319-17248-4_9}{\textbf{8966} 172--192}

\bibitem{gwpg.spec.org} SPEC/GWPG SPECviewperf 2020 v3.1 \url{https://gwpg.spec.org/benchmarks/benchmark/specviewperf-2020-v3-1/}

\bibitem{spec.org.accel2023} SPEC ACCEL 2023 \url{https://www.spec.org/accel2023/}

\bibitem{spec.org.cpu2017} SPEC CPU2017 \url{https://www.spec.org/cpu2017/}

\bibitem{QCion} Debnath S \textit{et al} 2016 Demonstration of a small programmable quantum computer with atomic qubits \textit{Nature} \href{https://doi.org/10.1038/nature18648}{\textbf{536} 63--66}

\bibitem{Zhang_2017} Zhang J \textit{et al} 2017 Observation of a many-body dynamical phase transition with a 53-qubit quantum simulator \textit{Nature} \href{https://doi.org/10.1038/nature24654}{\textbf{551} 601--604}

\bibitem{Arute2019} Arute F \textit{et al} 2019 Quantum supremacy using a programmable superconducting processor \textit{Nature} \href{http://dx.doi.org/10.1038/s41586-019-1666-5}{\textbf{574} 505--510}

\bibitem{ZHU2022240} Zhu Q \textit{et al} 2022 Quantum computational advantage via 60-qubit 24-cycle random circuit sampling \textit{Science} \href{https://doi.org/10.1016/j.scib.2021.10.017}{\textbf{67} 240--245}

\bibitem{PhysRevLett.127.180501} Wu Y \textit{et al} 2021 Strong quantum computational advantage using a superconducting quantum processor \textit{Phys. Rev. Lett.} \href{https://link.aps.org/doi/10.1103/PhysRevLett.127.180501}{\textbf{127} 180501}

\bibitem{QCPhoton} Zhong H \textit{et al} 2020 Quantum computational advantage using photons \textit{Science} \href{https://www.science.org/doi/10.1126/science.abe8770}{\textbf{370} 1460--1463}

\bibitem{PhysRevLett.127.180502} Zhong H, Deng Y, Qin J \textit{et al} 2021 Phase-programmable Gaussian boson sampling using stimulated squeezed light \textit{Phys. Rev. Lett.} \href{https://journals.aps.org/prl/abstract/10.1103/PhysRevLett.127.180502}{\textbf{127} 180502}

\bibitem{Bernien2017ProbingMD} Bernien H, Schwartz S, Keesling A \textit{et al} 2017 Probing many-body dynamics on a 51-atom quantum simulator \textit{Nature} \href{https://doi.org/10.1038/nature24622}{\textbf{551} 579--584}

\bibitem{Ebadi2020QuantumPO} Ebadi S \textit{et al} 2021 Quantum phases of matter on a 256-atom programmable quantum simulator \textit{Nature} \href{https://doi.org/10.1038/s41586-021-03582-4}{\textbf{595} 227--232}

\bibitem{farhi2000} Farhi E, Goldstone J, Gutmann S and Sipser M 2000 Quantum computation by adiabatic evolution (arXiv:\href{https://arxiv.org/abs/quant-ph/0001106}{quant-ph/0001106})

\bibitem{Kendon_2010} Kendon V M, Nemoto K and Munro W J 2010 Quantum analogue computing \textit{Philosophical Transactions of the Royal Society A} \href{http://doi.org/10.1098/rsta.2010.0017}{\textbf{368} 3609--3620}

\bibitem{doi:10.1126/science.273.5278.1073} Yao A C-C 1993 Quantum circuit complexity \textit{Proc. 34th Annual Symposium on Foundations of Computer Science (FOCS)} \href{https://ieeexplore.ieee.org/document/366852}{\textbf{1993} 352--361}

\bibitem{Raussendorf} Raussendorf R and Briegel H J 2001 A one-way quantum computer \textit{Phys. Rev. Lett.} \href{https://journals.aps.org/prl/abstract/10.1103/PhysRevLett.86.5188}{\textbf{86} 5188--5191}

\bibitem{LaRose2019overviewcomparison} LaRose R 2019 Overview and comparison of gate level quantum software platforms \textit{Quantum} \href{https://doi.org/10.22331/q-2019-03-25-130}{\textbf{3} 130}

\bibitem{Linke2017} Linke N M \textit{et al} 2017 Experimental comparison of two quantum computing architectures \textit{Proceedings of the National Academy of Sciences} \href{https://doi.org/10.1073/pnas.161802011}{\textbf{114} 3305--3310}

\bibitem{miessen2024benchmarkingdigitalquantumsimulations} Miessen A, Egger D, Tavernelli I and Mazzola G 2024 Benchmarking digital quantum simulations above hundreds of qubits using quantum critical dynamics (arXiv:\href{https://doi.org/10.48550/arXiv.2404.08053}{2404.08053})

\bibitem{Ilin_2024} Ilin Y and Arad I 2024 Learning a quantum channel from its steady-state \textit{New Journal of Physics} \href{http://dx.doi.org/10.1088/1367-2630/ad5464}{\textbf{26} 073003}

\bibitem{wright2024noisyintermediatescalequantumsimulation} Wright L \textit{et al} 2024 Noisy intermediate-scale quantum simulation of the one-dimensional wave equation (arXiv:\href{https://doi.org/10.48550/arXiv.2402.19247}{2402.19247})

\bibitem{Flannigan_2022} Flannigan S \textit{et al} 2022 Propagation of errors and quantitative quantum simulation with quantum advantage \textit{Quantum Sci. Technol.} \href{https://iopscience.iop.org/article/10.1088/2058-9565/ac88f5/meta}{\textbf{7} 045025}

\bibitem{Jnger2021} Jünger M \textit{et al} 2021 Quantum annealing versus digital computing: An experimental comparison \textit{ACM Journal} \href{https://doi.org/10.1145/3459606}{\textbf{26} 1084--6654}

\bibitem{havenstein2018comparisons} Havenstein C, Thomas D and Chandrasekaran S 2018 Comparisons of performance between quantum and classical machine learning \textit{SMU Data Science Review} \url{https://scholar.smu.edu/datasciencereview/vol1/iss4/11/}

\bibitem{Cugini2023} Cugini D \textit{et al} 2023 Comparing quantum and classical machine learning for vector boson scattering background reduction at the Large Hadron Collider \textit{Quantum Machine Intelligence} \href{http://dx.doi.org/10.1007/s42484-023-00106-3}{\textbf{5} 26}

\bibitem{kalai2022google} Kalai G, Rinott Y and Shoham T 2023 Google's quantum supremacy claim: Data, documentation, and discussion (arXiv:\href{https://doi.org/10.48550/arXiv.2210.12753}{2210.12753})

\bibitem{QuantumRam} Giovannetti V, Lloyd S and Maccone L 2008 Quantum random access memory \textit{Phys. Rev. Lett.} \href{http://dx.doi.org/10.1103/PhysRevLett.100.160501}{\textbf{100} 160501}

\bibitem{TROPP1980115} Tropp H 1980 The Smithsonian Computer History Project and some personal recollections \textit{Academic Press} \href{https://doi.org/10.1016/B978-0-12-491650-0.50016-2}{\textbf{Ch.} 115--130}

\bibitem{MooreLaw} Moore G E 2006 Cramming more components onto integrated circuits (reprinted from \textit{Electronics} 38(8), 1965) \textit{IEEE Solid-State Circuits Society Newsletter} \href{https://doi.org/10.1109/N-SSC.2006.4785860}{\textbf{11}(3) 33--35}

\bibitem{mythHarvard} Pawson R 2022 The myth of the Harvard architecture \textit{IEEE Annals of the History of Computing} \href{https://doi.org/10.1109/MAHC.2022.3175612}{\textbf{44}(4) 75--82}

\bibitem{Preskill2018} Preskill J 2018 Quantum computing in the NISQ era and beyond \textit{Quantum} \href{http://dx.doi.org/10.22331/q-2018-08-06-79}{\textbf{2} 79}

\bibitem{Fowler2012} Fowler A G, Mariantoni M, Martinis J M \textit{et al} 2012 Surface codes: Towards practical large-scale quantum computation \textit{Phys. Rev. A} \href{http://dx.doi.org/10.1103/PhysRevA.86.032324}{\textbf{86} 032324}

\bibitem{Andersen2020} Andersen C K \textit{et al} 2020 Repeated quantum error detection in a surface code \textit{Nature Physics} \href{http://dx.doi.org/10.1038/s41567-020-0920-y}{\textbf{16} 875--880}

\bibitem{Terhal2015} Terhal B M 2015 Quantum error correction for quantum memories \textit{Rev. Mod. Phys.} \href{http://dx.doi.org/10.1103/RevModPhys.87.307}{\textbf{87} 307--346}

\bibitem{suppressingQuantumErrors} Acharya R, Aleiner I, Allen R \textit{et al} 2023 Suppressing quantum errors by scaling a surface code logical qubit \textit{Nature} \href{http://dx.doi.org/10.1038/s41586-022-05434-1}{\textbf{614} 676--681}

\bibitem{10.1145/581771.581773} Ambainis A, Nayak A, Ta-Shma A and Vazirani U 2002 Dense quantum coding and quantum finite automata \textit{Journal of the ACM} \href{https://doi.org/10.1145/581771.581773}{\textbf{49} 496--511}

\bibitem{Maslov_2021} Maslov D \textit{et al} 2021 Quantum advantage for computations with limited space \textit{Nature Physics} \href{https://doi.org/10.1038/s41567-021-01271-7}{\textbf{17} 894--897}

\bibitem{PPP} (Note) There are notable exceptions, such as hybrid classical-quantum algorithms.


\bibitem{hashim2024} Hashim A, Nguyen L B, Goss N \textit{et al} 2024 A practical introduction to benchmarking and characterization of quantum computers (arXiv:\href{https://arxiv.org/abs/2408.12064}{2408.12064})

\bibitem{lall2025reviewcollectionmetricsbenchmarks}
D.~Lall, \textit{et al} 2025 A Review and Collection of Metrics and Benchmarks for Quantum Computers: definitions, methodologies and software (arXiv: \href{https://arxiv.org/abs/2502.06717}{2502.06717})

\bibitem{proctor2024benchmarkingquantumcomputers} Proctor T, Young K, Baczewski A and Blume-Kohout R 2024 Benchmarking quantum computers (arXiv:\href{https://doi.org/10.48550/arXiv.2407.08828}{2407.08828})

\bibitem{li2022qasmbench} Li A, Stein S, Krishnamoorthy S \textit{et al} 2020 QASMBench: A low-level QASM benchmark suite for NISQ evaluation and simulation (arXiv:\href{https://doi.org/10.48550/arXiv.2005.13018}{2005.13018})

\bibitem{PhysRevA.100.032328} Cross A, Bishop L, Sheldon S \textit{et al} 2019 Validating quantum computers using randomized model circuits \textit{Phys. Rev. A} \href{https://doi.org/10.1103/PhysRevA.100.032328}{\textbf{100} 032328}

\bibitem{QuantumLINPACK} Dong Y and Lin L 2021 Random circuit block-encoded matrix and a proposal of quantum LINPACK benchmark \textit{Phys. Rev. A} \href{https://doi.org/10.1103/PhysRevA.103.062412}{\textbf{103} 062412}

\bibitem{Martiel2021BenchmarkingQC} Martiel S, Ayral T and Allouche C 2021 Benchmarking quantum coprocessors in an application-centric, hardware-agnostic, and scalable way \textit{IEEE Trans. Quantum Engineering} \href{https://doi.org/10.1109/TQE.2021.309020}{\textbf{2} 3102011}

\bibitem{Gheorghiu2018} Gheorghiu A, Kapourniotis T and Kashefi E 2018 Verification of quantum computation: An overview of existing approaches \textit{Theory of Computing Systems} \href{https://doi.org/10.1007/s00224-018-9872-3}{\textbf{63} 715--808}

\bibitem{AaronsonPrize} Aaronson S 2006 The $25,000$ prize for quantum-supremacy experiments \url{https://scottaaronson.blog/?p=284}

\bibitem{Mahadev2018} Mahadev U 2018 Classical verification of quantum computations \textit{Proc. 59th IEEE Symposium on Foundations of Computer Science (FOCS)} \href{http://dx.doi.org/10.1109/FOCS.2018.00033}{\textbf{2018} 259--267}

\bibitem{Proctor2021} Proctor T \textit{et al} 2022 Measuring the capabilities of quantum computers \textit{Nature Physics} \href{https://doi.org/10.1038/s41567-021-01409-7}{\textbf{18} 75--79}



\bibitem{PhysRevA.40.2847} Vogel K and Risken H 1989 Determination of quasiprobability distributions in terms of probability distributions for the rotated quadrature phase \textit{Phys. Rev. A} \href{https://link.aps.org/doi/10.1103/PhysRevA.40.2847}{\textbf{40} 2847--2849}

\bibitem{FormalVerificationQuantumPrograms} Lewis M, Soudjani S and Zuliani P 2023 Formal verification of quantum programs: Theory, tools and challenges \textit{ACM Transactions on Quantum Computing} \href{https://doi.org/10.1145/3624483}{\textbf{5}(1) 1--35}

\bibitem{GarcadelaBarrera2021} García de la Barrera A \textit{et al} 2021 Quantum software testing: State of the art \textit{Journal of Software: Evolution and Process} \href{https://doi.org/10.1002/smr.2419}{\textbf{35} e2419}

\bibitem{8805685} Miranskyy A and Zhang L 2019 On testing quantum programs \textit{Proc. IEEE/ACM 41st International Conference on Software Engineering: New Ideas and Emerging Results (ICSE-NIER)} \href{https://doi.org/10.1109/ICSE-NIER.2019.00023}{\textbf{2019} 57--60}

\bibitem{Blume_Kohout_2020} Blume-Kohout R and Young K C 2020 A volumetric framework for quantum computer benchmarks \textit{Quantum} \href{https://doi.org/10.22331/q-2020-11-15-362}{\textbf{4} 362}

\bibitem{eisert2020quantum} Eisert J, Hangleiter D, Walk N \textit{et al} 2020 Quantum certification and benchmarking \textit{Nature Reviews Physics} \href{https://www.nature.com/articles/s42254-020-0186-4}{\textbf{2} 382--390}

\bibitem{SoKBenchmarkingQuantumComputer} Wang J, Guo G and Shan Z 2022 SoK: Benchmarking the performance of a quantum computer \textit{Entropy} \href{https://doi.org/10.3390/e24101467}{\textbf{24}(10) 1467}

\bibitem{https://doi.org/10.48550/arxiv.2110.14108} Wack A \textit{et al} 2021 Quality, speed, and scale: Three key attributes to measure the performance of near-term quantum computers (arXiv:\href{https://doi.org/10.48550/arXiv.2110.14108}{2110.14108})


\bibitem{https://doi.org/10.48550/arxiv.1912.00546} Resch S and Karpuzcu U 2019 Benchmarking quantum computers and the impact of quantum noise (arXiv:\href{https://doi.org/10.48550/arXiv.1912.00546}{1912.00546})

\bibitem{quantum_consortium} Quantum Economic Development Consortium \url{https://quantumconsortium.org/}

\bibitem{lubinski2023applicationoriented} Lubinski T \textit{et al} 2021 Application-oriented performance benchmarks for quantum computing (arXiv:\href{https://doi.org/10.48550/arXiv.2110.03137}{2110.03137})

\bibitem{Mills2021} Mills D \textit{et al} 2021 Application-motivated, holistic benchmarking of a full quantum computing stack \textit{Quantum} \href{https://doi.org/10.22331/q-2021-03-22-415}{\textbf{5} 415}

\bibitem{Finzgar_2022} Finžgar J \textit{et al} 2022 QUARK: A framework for quantum computing application benchmarking (arXiv:\href{https://doi.org/10.48550/arXiv.2202.03028}{2202.03028})

\bibitem{PhysRevLett.129.150502} Proctor T, Seritan S, Rudinger K, Nielsen E, Blume-Kohout R and Young K 2022 Scalable randomized benchmarking of quantum computers using mirror circuits \textit{Phys. Rev. Lett.} \href{https://doi.org/10.1103/PhysRevLett.129.150502}{\textbf{129} 150502}


\bibitem{doi:10.1137/18M120275X} Aaronson S 2020 Shadow tomography of quantum states \textit{SIAM Journal on Computing} \href{https://epubs.siam.org/doi/10.1137/18M120275X}{\textbf{49}(5) 1030--1074}

\bibitem{article12} Huang H, Kueng R and Preskill J 2020 Predicting many properties of a quantum system from very few measurements \textit{Nature Physics} \href{https://www.nature.com/articles/s41567-020-0932-7}{\textbf{16} 1050--1057}

\bibitem{zhou2023efficient} Zhou T and Zhang P 2021 Efficient classical shadow tomography through many-body localization dynamics (arXiv:\href{https://doi.org/10.48550/arXiv.2309.01258}{2309.01258})

\bibitem{10.5555/863284} Kitaev A Yu, Shen A H and Vyalyi M N 2002 Classical and quantum computation \textit{American Mathematical Society}

\bibitem{nielsen00} Nielsen M and Chuang I 2010 Quantum computation and quantum information \textit{Cambridge University Press}

\bibitem{Emerson_2005} Emerson J, Alicki R and Życzkowski K 2005 Scalable noise estimation with random unitary operators \textit{Journal of Optics B: Quantum and Semiclassical Optics} \href{https://doi.org/10.1088/1464-4266/7/10/021}{\textbf{7}(10) S347--S352}

\bibitem{Dankert_2009} Dankert C \textit{et al} 2009 Exact and approximate unitary 2-designs and their application to fidelity estimation \textit{Phys. Rev. A} \href{http://dx.doi.org/10.1103/PhysRevA.80.012304}{\textbf{80} 012304}

\bibitem{PhysRevLett.106.180504} Magesan E, Gambetta J M and Emerson J 2011 Scalable and robust randomized benchmarking of quantum processes \textit{Phys. Rev. Lett.} \href{https://link.aps.org/doi/10.1103/PhysRevLett.106.180504}{\textbf{106} 180504}

\bibitem{Wallman_2015} Wallman J \textit{et al} 2015 Estimating the coherence of noise \textit{New Journal of Physics} \href{http://dx.doi.org/10.1088/1367-2630/17/11/113020}{\textbf{17} 113020}

\bibitem{Magesan_2012} Magesan E \textit{et al} 2012 Efficient measurement of quantum gate error by interleaved randomized benchmarking \textit{Phys. Rev. Lett.} \href{http://dx.doi.org/10.1103/PhysRevLett.109.080505}{\textbf{109} 080505}

\bibitem{Hines2024} Hines J, Hothem D, Blume-Kohout R, Whaley B and Proctor T 2024 Fully scalable randomized benchmarking without motion reversal \textit{PRX Quantum} \href{https://doi.org/10.1103/PRXQuantum.5.030334}{\textbf{5} 030334}

\bibitem{polloreno2023} Polloreno A M, Carignan-Dugas A, Hines J, Blume-Kohout R, Young K and Proctor T 2023 A theory of direct randomized benchmarking (arXiv:\href{https://arxiv.org/abs/2302.13853}{2302.13853})


\bibitem{Holmes_2020} Holmes A, Johri S, Guerreschi G G \textit{et al} 2020 Impact of qubit connectivity on quantum algorithm performance \textit{Quantum Sci. Technol.} \href{https://iopscience.iop.org/article/10.1088/2058-9565/ab73e0}{\textbf{5}(2) 025009}

\bibitem{PRXQuantum.5.040326} Swiadek F, Shillito R, Magnard P \textit{et al} 2024 Enhancing dispersive readout of superconducting qubits through dynamic control of the dispersive shift: Experiment and theory \textit{PRX Quantum} \href{https://doi.org/10.1103/PRXQuantum.5.040326}{\textbf{5} 040326}

\bibitem{PhysRevApplied.23.054057} Chatterjee A, Schwinger J and Gao Y Y 2025 Enhanced qubit readout via reinforcement learning \textit{Phys. Rev. Applied} \href{https://doi.org/10.1103/PhysRevApplied.23.054057}{\textbf{23} 054057}

\bibitem{Seif_2018} Seif A, Landsman K A, Linke N M, Figgatt C, Monroe C and Hafezi M 2018 Machine learning assisted readout of trapped-ion qubits \textit{J. Phys. B: At. Mol. Opt. Phys.} \href{https://doi.org/10.1088/1361-6455/aad62b}{\textbf{51}(17) 174006}

\bibitem{Viola1998} Viola L and Lloyd S 1998 Dynamical suppression of decoherence in two-state quantum systems \textit{Phys. Rev. A} \href{https://doi.org/10.1103/PhysRevA.58.2733}{\textbf{58}(4) 2733--2744}

\bibitem{van_der_Schoot_2023} van der Schoot W, Wezeman R, Eendebak P \textit{et al} 2023 Evaluating three levels of quantum metrics on quantum-inspire hardware \textit{Quantum Information Processing} \href{https://link.springer.com/article/10.1007/s11128-023-04184-x}{\textbf{22} 451}

\bibitem{Werninghaus2021} Werninghaus M \textit{et al} 2021 Leakage reduction in fast superconducting qubit gates via optimal control \textit{npj Quantum Information} \href{https://doi.org/10.1038/s41534-020-00346-2}{\textbf{7} 14}

\bibitem{Yan2018} Yan F \textit{et al} 2018 Tunable coupling scheme for implementing high-fidelity two-qubit gates \textit{Phys. Rev. Applied} \href{https://doi.org/10.1103/PhysRevApplied.10.054062}{\textbf{10} 054062}

\bibitem{mckay2023benchmarkingquantumprocessorperformance} McKay D \textit{et al} 2023 Benchmarking quantum processor performance at scale (arXiv:\href{https://doi.org/10.48550/arXiv.2311.05933}{2311.05933})

\bibitem{9459509} Martiel S, Ayral T and Allouche C 2021 Benchmarking quantum coprocessors in an application-centric, hardware-agnostic, and scalable way \textit{IEEE Trans. Quantum Engineering} \href{https://ieeexplore.ieee.org/document/9459509}{\textbf{2} 3102011}

\bibitem{doi:10.1126/sciadv.aau0823} Hamerly R \textit{et al} 2019 Experimental investigation of performance differences between coherent Ising machines and a quantum annealer \textit{Science Advances} \href{https://www.science.org/doi/10.1126/sciadv.aau0823}{\textbf{5}(5) eaau0823}

\bibitem{vanderschoot2023qscore} van der Schoot W \textit{et al} 2023 Q-score Max-Clique: The first quantum metric evaluation on multiple computational paradigms (arXiv:\href{https://doi.org/10.48550/arXiv.2302.00639}{2302.00639})

\bibitem{wack2021quality} Wack A \textit{et al} 2021 Quality, speed, and scale: Three key attributes to measure the performance of near-term quantum computers (arXiv:\href{https://doi.org/10.48550/arXiv.2110.14108}{2110.14108})

\bibitem{www.ibm.com} IBM Quantum blog: Quantum metric—layer fidelity \url{https://www.ibm.com/quantum/blog/quantum-metric-layer-fidelity}

\bibitem{chen2023benchmarking} Chen J, Nielsen E, Ebert M \textit{et al} 2023 Benchmarking a trapped-ion quantum computer with 29 algorithmic qubits (arXiv:\href{ https://doi.org/10.48550/arXiv.2308.05071}{2308.05071})

\bibitem{Qsupremacy} Boixo S, Isakov S, Smelyanskiy V \textit{et al} 2018 Characterizing quantum supremacy in near-term devices \textit{Nature Physics} \href{ https://doi.org/10.1038/s41567-018-0124-x}{\textbf{14} 595--600}

\bibitem{e24020244} Dasgupta S and Humble T S 2022 Characterizing the reproducibility of noisy quantum circuits \textit{Entropy} \href{ https://doi.org/10.3390/e24020244}{\textbf{24}(2) 244}

\bibitem{10.5555/3135595.3135617} Aaronson S and Chen L 2016 Complexity-theoretic foundations of quantum supremacy experiments (arXiv:\href{https://doi.org/10.48550/arXiv.1612.05903}{1612.05903})

\bibitem{Mills_2021} Mills D, Sivarajah S, Scholten T L and Duncan R 2021 Application-motivated, holistic benchmarking of a full quantum computing stack \textit{Quantum} \href{http://dx.doi.org/10.22331/q-2021-03-22-415}{\textbf{5} 415}

\bibitem{mari2023} Mari A 2023 Counting collisions in random circuit sampling for benchmarking quantum computers \textit{AIP Advances} \href{https://doi.org/10.1063/5.0219266}{\textbf{13} 075132}

\bibitem{QPACK2021} Mesman K, Al-Ars Z and Möller M 2021 QPack: Quantum approximate optimization algorithms as universal benchmark for quantum computers (arXiv:\href{https://doi.org/10.48550/arXiv.2103.17193}{2103.17193})

\bibitem{QPACK_} Donkers H \textit{et al} 2022 QPack scores: Quantitative performance metrics for application-oriented quantum computer benchmarking (arXiv:\href{https://doi.org/10.48550/arXiv.2205.12142}{2205.12142})

\bibitem{QuantumAlgorithms} Montanaro A 2015 Quantum algorithms: An overview \textit{npj Quantum Information} \href{https://doi.org/10.1038/npjqi.2015.23}{\textbf{2} 15023}

\bibitem{9773202} Tomesh T \textit{et al} 2022 SupermarQ: A scalable quantum benchmark suite (arXiv:\href{https://doi.org/10.48550/arXiv.2202.11045}{2202.11045})

\bibitem{QuantumChemistry} McCaskey A J, Parks Z P, Jakowski J \textit{et al} 2019 Quantum chemistry as a benchmark for near-term quantum computers \textit{npj Quantum Information} \href{https://doi.org/10.1038/s41534-019-0209-0}{\textbf{5} 99}

\bibitem{VQE-Photonic} Peruzzo A \textit{et al} 2014 A variational eigenvalue solver on a quantum processor \textit{Nature Communications} \href{https://doi.org/10.1038/ncomms5213}{\textbf{5} 4213}

\bibitem{PhysRevX.6.031007} O'Malley P J J, Babbush R, Kivlichan I D \textit{et al} 2016 Scalable quantum simulation of molecular energies \textit{Phys. Rev. X} \href{https://doi.org/10.1103/PhysRevX.6.031007}{\textbf{6} 031007}

\bibitem{PhysRevX.8.031022} Hempel C \textit{et al} 2018 Quantum chemistry calculations on a trapped-ion quantum simulator \textit{Phys. Rev. X} \href{https://doi.org/10.1103/PhysRevX.8.031022}{\textbf{8} 031022}


\bibitem{SPECHPGBenchmarksHistory} Eigenmann R 2001 Performance evaluation and benchmarking with realistic applications \textit{The MIT Press}

\bibitem{Neill_2018} Neill C, Roushan P, Kechedzhi K \textit{et al} 2018 A blueprint for demonstrating quantum supremacy with superconducting qubits \textit{Science} \href{https://www.science.org/doi/10.1126/science.aao4309}{\textbf{360}(6385) 195--199}

\bibitem{aharonov2012quantum} Aharonov D and Vazirani U 2012 Is quantum mechanics falsifiable? A computational perspective on the foundations of quantum mechanics (arXiv:\href{https://doi.org/10.48550/arXiv.1206.3686}{1206.3686})

\bibitem{cross2017open} Cross A \textit{et al} 2017 Open Quantum Assembly Language (arXiv:\href{https://doi.org/10.48550/arXiv.1707.03429}{1707.03429})

\bibitem{1580386} Svore K M \textit{et al} 2006 A layered software architecture for quantum computing design tools \textit{Computer} \href{https://doi.org/10.1109/MC.2006.4}{\textbf{39} 74--83}

\bibitem{9951320} Younis E and Iancu C 2022 Quantum circuit optimization and transpilation via parameterized circuit instantiation \textit{IEEE International Conference on Quantum Computing and Engineering (QCE)} \href{https://doi.ieeecomputersociety.org/10.1109/QCE53715.2022.00068}{\textbf{2022} 465--475}

\bibitem{Hennessy2011-kt} Hennessy J and Patterson D 2017 Computer architecture 6th ed.

\bibitem{BenchmarkingOptimization} Bartz-Beielstein T \textit{et al} 2020 Benchmarking in optimization: Best practice and open issues (arXiv:\href{https://doi.org/10.48550/arXiv.2007.03488}{2007.03488})

\bibitem{TestingHeuristics} Hooker J 1995 Testing heuristics: We have it all wrong \textit{Journal of Heuristics} \href{https://doi.org/10.1007/BF02430364}{\textbf{1} 33--42}





\end{thebibliography}
\end{document}